%% file: ms.tex
\documentclass[12pt,preprint]{aastex}

\shorttitle{A universal GRB energy-luminosity relation}
\shortauthors{Willingale et al.}

\begin{document}


\title{A universal GRB photon energy-peak luminosity relation}


\author{R. Willingale\altaffilmark{1},
P.T. O'Brien\altaffilmark{1},
M.R. Goad\altaffilmark{1},
J.P. Osborne\altaffilmark{1},
K.L. Page\altaffilmark{1},
N.R. Tanvir\altaffilmark{1}
}


\altaffiltext{1}{Department of Physics and Astronomy, University of Leicester,
LE1 7RH, UK}


\begin{abstract}
The energetics and emission mechanism of GRBs are not well understood.
Here we demonstrate that the instantaneous peak flux or equivalent isotropic
peak luminosity, $L_{iso}$ ergs s$^{-1}$, rather than the integrated
fluence or equivalent isotropic energy, $E_{iso}$ ergs,
underpins the known high-energy correlations.
Using new spectral/temporal parameters calculated for
101 bursts with redshifts from {\em BATSE}, {\em BeppoSAX}, {\em HETE-II}
and {\em Swift} we describe a parameter space which
characterises the apparently diverse properties of the prompt emission.
We show that a source frame characteristic-photon-energy/peak luminosity
ratio, $K_{z}$, can be constructed
which is constant within a factor of 2
for all bursts whatever their duration, spectrum,
luminosity and the instrumentation used to detect them.
The new parameterization embodies the Amati relation but
indicates that some correlation between $E_{peak}$ and $E_{iso}$ follows
as a direct mathematical inference from the Band function and that
a simple transformation of $E_{iso}$ to $L_{iso}$
yields a universal high energy correlation for GRBs.
The existence of $K_{z}$ indicates that the mechanism responsible
for the prompt emission from all GRBs is probably predominantly thermal.
\end{abstract}





\section{Introduction}

The energetics of the central engine which powers the explosion
responsible for a GRB are both intriguing and fundamental to our
understanding of these cosmic events. The isotropic energy outflow at source,
estimated using the integrated gamma-ray fluence, is enormous, up to
$E_{iso}\sim10^{54}$
ergs, and even if the outflow is collimated in jets the total
energy involved is still huge,
$E_{\gamma}\sim10^{51}$ ergs. The possibility that the explosion taps
a standard energy resevoir has been pursued by many authors following
the initial suggestion from Frail et al. (2001). If this total energy
available were, indeed, roughly constant (or predictable through other means)
and we could reliably estimate the collimation,
then GRBs could be used as a cosmological probe to very high redshifts,
Bloom et al. (2003), Ghirlanda et al. (2004).

Early on it was noted that, based on analysis of {\em BATSE} data,
there was a correlation between $E_{p}$, the peak of
$E.F(E)$ where $F(E)$ ergs cm$^{-2}$ keV$^{-1}$ is the observed
spectrum, and the fluence (Mallozzi et al. 1995,
Lloyd et al. 2000). When redshifts
became available for long bursts
the isotropic energy, $E_{iso}$, could be estimated from 
the fluence and the peak energy could be transformed into the source
frame, $E_{pz}$, the so-called Amati relation, a correlation
between $E_{iso}$ and $E_{pz}$ in the sense that more energetic
bursts have a higher $E_{pz}$, was discovered using data from
{\em BeppoSAX}, (Amati et al. 2002). This correlation has subsequently been
confirmed and extended although there remain many significant
outliers, including all short bursts.
The physical origin of the correlation may be associated
with the emission mechanisms operating in the fireball but the theoretical
details are far from settled
(see the discussion by Amati (2006) and references therein).
More recently a tighter correlation between $E_{iso}$, $E_{pz}$
and the jet break time, $t_{break}$, measured in the optical afterglow
has been reported (Ghirlanda et al. 2004).
This is explained in terms of a modification to the Amati relation in which
$E_{iso}$ is corrected to a true collimated energy, $E_{\gamma}$,
using an estimate of the collimation angle derived from $t_{break}$.
The details of the 
collimation correction depend on the density and density profile
of the circumburst medium, Nava et al. (2006) and references therein.
Multivariable regression analysis was performed by Liang \& Zhang (2005)
to derive a model-independent relationship,
$E_{iso}\propto E_{pz}^{1.94}t_{zbreak}^{-1.24}$, indicating that the
rest-frame break time of the optical afterglow, $t_{zbreak}$
was indeed correlated with the prompt emission parameters.

Other studies have concentrated on the properties of the
isotropic peak (maximum) luminosity, $L_{iso}$
ergs s$^{-1}$, measured over some short time scale $\approx1$ s,
rather than the time integrated isotropic energy, $E_{iso}$.
Yonetoku et al. (2004) noted a correlation between $L_{iso}$ and
$E_{pz}$ for 16 GRBs with firm redshifts. A correlation
between $L_{iso}$ and the spectral lag was first identified by Norris et al.
(2000) and explained in terms of the evolution of $E_{peak}$ with time.
The shocked material responsible for
the gamma-ray emission is expected to cool at a rate proportional
to the gamma-ray luminosity and it has been suggested that
$E_{peak}$ traces the cooling (Schaefer 2004). 
A similar
correlation between $L_{iso}$ and the variability of the GRB ($V$) was
described by Reichart et al. (2001). The origin of the $L_{iso}-V$ 
relation is likely to be related to the physics of the relativistic shocks
and the bulk Lorentz factor of the outflow.
It could be that 
high $\Gamma_{outflow}$ results in high $L_{iso}$ and $V$ while
lower luminosity and variability are expected if $\Gamma_{outflow}$ is low
(see, for example, M\'{e}sz\'{a}ros et al. 2002).
A rather bizzare correlation involving $L_{iso}$, $E_{pz}$ and variability
was found by Firmani et al. (2006). They employed the ``high signal'' time,
$T_{45}$, as formulated by Reichart et al. (2001) in their study of
variability, and showed that $L_{iso}\propto E_{pz}^{1.62}T_{45}^{-0.49}$
for 19 GRBs with a spread much narrower than that of the Amati relation.
There is currently no explanation for such a correlation although
it may be connected with the spectral lag and variability
correlations and the Amati relation.

The correlation between $E_{iso}$ and $E_{pz}$ supplemented by additional
empirical information can be
used in pseudo redshift indicators, for example Atteia (2003), 
Pelangeon \& Atteia (2006), but the intrinsic spread in the correlation
and uncertainty about the underlying physical interpretation
introduce errors, typically
of a factor $\sim2$. It may be possible to reduce the errors by
simultaneous application of several independent luminosity/energy 
correlations, and extension of the Hubble Diagram to high redshifts
using GRBs has been attempted, see for example Schaefer (2007).
However, it is not clear that the correlations briefly described above
are truly independent and there may be some underlying principle or
mechanism which connects them all together. Recently, and more controversially,
Butler et al. (2007) have raised serious doubts about the validity
of these correlations suggesting that it is likely that they
are introduced by observational/instrumental bias and have nothing to
do with the physical properties of the GRBs and hence they conclude
that GRBs are probably
useless as cosmological probes. Here we take a new look at the source
frame spectral and temporal properties of a large number of GRBs for
which we have redshifts in order
to try and understand what really correlates with what and whether or
not this can provide useful intrinsic information about the GRBs and
what drives them. In this analysis we include the short-duration
GRBs which may share a similar emission mechanism with long bursts
despite probably having different progenitors.

\section{Source frame spectra of the prompt emission}

The profile of the prompt energy spectrum of all GRBs is well represented by
a Band function (Band et al. 1993),
\[
B(E)=E^{-\beta_{X}} \exp(-E/E_{c}),\:\: E \le E_{c}(\beta_{\gamma}-\beta_{X})
\]
\begin{equation}
B(E)=E^{-\beta_{\gamma}} \exp(\beta_{\gamma}-\beta_{X})
[(\beta_{\gamma}-\beta_{X})E_{c}]^{\beta_{\gamma}-\beta_{X}},\:\:
E \ge E_{c}(\beta_{\gamma}-\beta_{X})
\label{eq1}
\end{equation}
where $\beta_{X}$ and $\beta_{\gamma}$ are the spectral power law indices
at low (X-ray) and high ($\gamma$-ray)
energies respectively and $E_{c}$ keV is the high cut-off energy. Note
that in the original formulation of Band et al. (1993) photon indices were
used and the profile described the photon number density
(because these are the parameters which most closely describe the detected
count spectrum which is fitted). Here we choose to
use an energy density profile and energy spectral indices.
The observed total fluence is
\begin{equation}
F_{tot}=\int_{E_{1}}^{E_{2}}F(E)dE=
N_{tot}\int_{E_{1}}^{E_{2}}B(E)dE
\label{eq2}
\end{equation}
ergs cm$^{-2}$, where $N_{tot}$ is the normalisation in
ergs cm$^{-2}$ keV$^{-1}$ at 1 keV and $E_{1}$ to $E_{2}$ is the
observed energy band. Spectral fitting of the observed count spectrum
will yield values for $\beta_{X}$, $\beta_{\gamma}$, $E_{c}$
and $N_{tot}$. The cut-off energy, $E_{c}$, is often converted
to the peak energy of the $E.F(E)$
spectrum which is given by $E_{p}=(1-\beta_{X})E_{c}$ and the
normalisation may be expressed as the fluence, $F_{tot}$, rather than the energy
density at 1 keV, $N_{tot}$. However, the separation of the fluence into a
normalisation term and a spectral integral is central to the development
of the argument which follows.
Table \ref{tab1} gives the spectral
parameters for 101 GRBs for which we have redshift values and a prompt
light curve.
The spectral
parameters for bursts detected by {\em BATSE}, {\em BeppoSAX}, {\em HETE-2}
and {\em Konus/WIND}
were taken from the references cited. The values for {\em Swift} bursts
were derived from the {\em BAT} spectra supplemented by detections
by {\em INTEGRAL} and {\em Konus/WIND} where available. Many of the
{\em Swift} spectra ($\approx 40$) are adequately fitted by a simple
power law or a cut-off power law with $E_{c}$ fixed. For these
bursts a cut-off power law model was used with $E_{c}=150$ keV
(corresponding to the upper limit of the BAT energy band). Providing
the fitted $\beta_{X}<1$ the fitted function has a peak in $E.F(E)$
and a value for the peak energy can then be estimated.
The spectra of 7 very soft {\em Swift} bursts with redshifts
(GRB050406, GRB050416A, GRB050824, GRB051016B, GRB060512, GRB060926 and
GRB070419A)
gave $\beta_{X}\ge 1$ and these were discarded because, for such spectra,
we have no meaningful estimate of $E_{p}$.
Such GRBs are normally designated as X-ray flashes (XRFs) and
the exceptionally high $\beta_{X}$ values may arise because we are actually
observing the high energy tail ($\beta_{\gamma}$) and not the lower energy
power law in the Band function. Alternatively it may be that such
soft spectra are the result of a second soft X-ray component which
dominates in these objects.

The equivalent isotropic energy from the source is given by
\begin{equation}
E_{iso}=\frac{4\pi d_{L}^{2} N_{tot}}{(1+z)^{2-\beta_{X}}}
I_{bol}(E_{pz},\beta_{X},\beta_{\gamma})
\label{eq3}
\end{equation}
ergs, where $d_{L}$ is the luminosity distance corresponding to the
redshift $z$ under some cosmology,
$I_{bol}(E_{pz},\beta_{X},\beta_{\gamma})$
is the bolometric integral of the spectral energy profile in the source
frame, $B_{z}$, taken over the wide energy band 1 keV to 10 MeV
\begin{equation}
I_{bol}(E_{pz},\beta_{X},\beta_{\gamma})=\int_{1}^{10^{4}} B_{z}(E)dE
\label{eq4}
\end{equation}
and $E_{pz}=E_{p}(1+z)$ is the peak energy in the source frame.
The first term in Equation \ref{eq3}
is the equivalent isotropic energy density, $Q_{z}$
ergs keV$^{-1}$ at 1 keV in the source frame.
\begin{equation}
Q_{z}=\frac{4\pi d_{L}^{2} N_{tot}}{(1+z)^{2-\beta_{X}}}.
\label{eq5}
\end{equation}
A factor $1/(1+z)^{1-\beta_{X}}$ arises because we have shifted the
normalisation from 1 keV in observer frame to 1 keV in the source frame.
The remaining factor of $1/(1+z)$ accounts for the time-dilation of the
duration over which the bursts are seen.
It is pertinent to transform this to the isotropic energy density at the peak
energy, $E_{pz}$ keV, in the source frame,
\begin{equation}
Q_{pz}=Q_{z}\exp[(1-\beta_{X})(E_{pz}^{-1}-1)] E_{pz}^{-\beta_{X}}
\label{eq6}
\end{equation}
ergs keV$^{-1}$ so that
the spectrum normalisation is specified at a characteristic
energy in or close to the
observed $\gamma$-ray energy band.
We can then write Equation \ref{eq3} as
\begin{equation}
E_{iso}=Q_{pz}E_{wz},
\label{eq7}
\end{equation}
where
\begin{equation}
E_{wz}=
\exp [(\beta_{X}-1)(E_{pz}^{-1}-1)] 
E_{pz}^{\beta_{X}}
I_{bol}(E_{pz},\beta_{X},\beta_{\gamma})
\label{eq8}
\end{equation}
keV is a {\em characteristic photon energy} which depends on the profile
of the energy spectrum and the limits adopted for the integration
and it serves to convert from an energy density ($Q_{pz}$ ergs keV$^{-1}$)
at the peak of the $E.F_{z}(E)$
spectrum to the total isotropic energy ($E_{iso}$ ergs).
The isotropic energy spectrum in the source frame is given  by
\begin{equation}
F_{z}(E)=Q_{z}B_{z}(E)=
Q_{pz} \exp [(\beta_{X}-1)(E_{pz}^{-1}-1)]
E_{pz}^{\beta_{X}}B_{z}(E)
\label{eq9}
\end{equation}
ergs keV$^{-1}$.
The source frame spectra of the GRBs listed in Table \ref{tab1}
are shown in Figure \ref{fig1} with the spectral energy density $Q_{pz}$ marked
at energy $E_{pz}$ keV.
In the majority of spectra the high energy spectral index is not measured but
set to $\beta_{\gamma}=1.3$ which is the approximate average
found by {\em BATSE}. 
Figure \ref{fig2} shows the corresponding
$E.F_{z}(E)$ spectra in ergs.
We assumed a cosmology with $H_{0}=71$ km s$^{-1}$
Mpc$^{-1}$, $\Lambda=0.27$ and $\Omega=0.73$ to calculate the 
luminosity distance $d_{L}$.

\section{The Amati relation}

The Amati relation is a correlation between $E_{pz}$ and $E_{iso}$,
first reported by Amati et al. (2002), and subsequently shown to be obeyed
by the majority of long GRBs although there is a fairly large scatter.
The top left panel of Figure \ref{fig3} shows the histogram of
isotropic energy values, $E_{iso}$, calculated using Equation \ref{eq7} using
the spectral parameters in Table \ref{tab1} and redshift in Table \ref{tab3}.
A large range of values for $E_{iso}$ is produced because
of the spread in the isotropic
energy density at the peak $Q_{pz}$, the peak energy $E_{pz}$
and the bolometric integral $I_{bol}$. The top right panel of
Figure \ref{fig3} shows the peak energy values, $E_{pz}$, plotted
against the characteristic energy, $E_{wz}$. There is a tight correlation
between these 2 parameters because of the form of the Band function.
To a first approximation $E_{pz}=0.23 E_{wz}$ (the solid line
in Figure \ref{fig3}) although
the best fit correlation is a little steeper ($E_{pz}\propto E_{wz}^{1.14}$)
and the small scatter
evident in Figure \ref{fig3} is introduced by differences in the spectral
indices, $\beta_{X}$ and $\beta_{\gamma}$.
In fact, the bolometric integral is well approximated by a function of the form
\begin{equation}
I_{bol}\approx I_{fit}=E_{pz}^{1+c_{1}}
\exp(c_{0}+c_{2}\beta_{X}+c_{3}\beta_{\gamma})
\label{eq10}
\end{equation}
where the coefficients $c_{0},c_{1},c_{2},c_{3}$ can be found by a
least squares fitting procedure. A comparison of $I_{fit}$ and $I_{bol}$
for the GRBs listed in Table \ref{tab1} is shown in the bottom left panel of
Figure \ref{fig3} together with the best fit coefficients.
We can use $I_{fit}$ in place of $I_{bol}$ and
estimate $E_{iso}\approx E_{fit}$. The distribution of the
ratio $E_{iso}/E_{fit}$ is
shown in the bottom right-hand panel of Figure \ref{fig3}. For the majority
of objects the estimation, $E_{fit}$, is within $\pm10\%$ of the value obtained
by numerical integration. There are a few GRBs with a larger discrepancy
but all are within $\pm20\%$ which is a very small perturbation in comparison
with the dynamic range of the $E_{iso}$ values.

Using $I_{bol}\approx I_{fit}$  we can express $E_{iso}$ as an explicit
function of $E_{pz}$:
\begin{equation}
E_{iso}\approx Q_{pz} E_{pz}^{\beta_{X}+1+c_{1}}
\exp(1+c_{0}+(c_{2}-1)\beta_{X}+c_{3}\beta_{\gamma}).
\label{eq11}
\end{equation}
The immediate origin of the Amati relationship is now clear.
Given Equation \ref{eq11} some degree of correlation between $E_{pz}$ and
$E_{iso}$ is guaranteed. The nature and spread of this correlation
will depend on the relationship between
the flux density, $Q_{pz}$, and the peak energy, $E_{pz}$, and the distribution
of spectral index $\beta_{X}$. It could be that $Q_{pz}$ and $E_{pz}$ are
correlated in such a way to cancel the apparent dependence on $E_{pz}$ but
this is highly unlikely.
This correlation arises because the GRB spectral profile has the form of 
Band function (Equation \ref{eq1}) with a particular range of values
for the spectral indices, $\beta_{X}$, $\beta_{\gamma}$,
and the energy $E_{c}$.
So understanding where the Amati
relation comes from is really the same as understanding why the spectra
have this functional form in the first place.

\begin{figure}[!htp]
\begin{center}
\includegraphics[height=15cm,angle=-90]{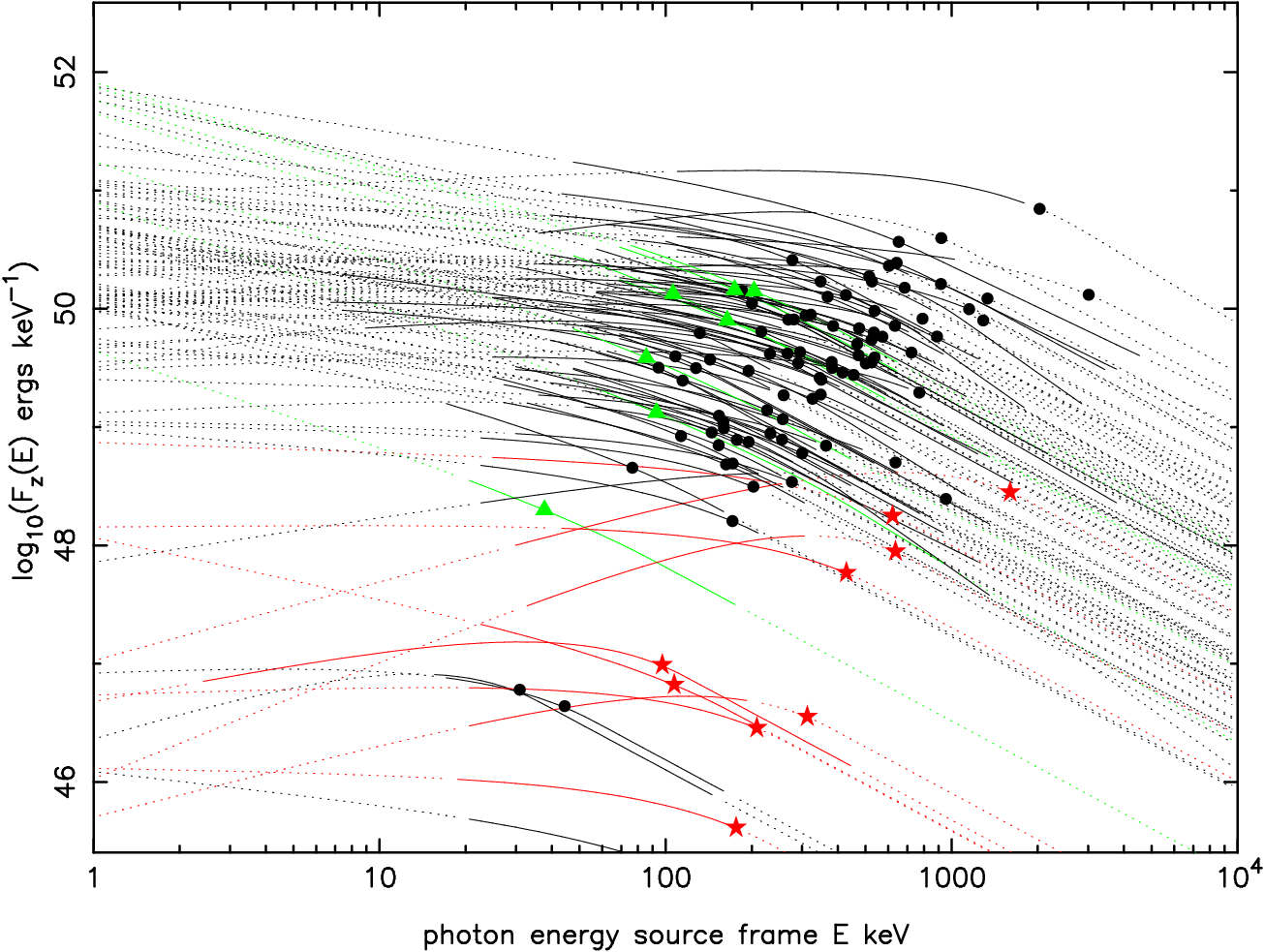}
\end{center}
\caption{The 1 keV to 10 MeV
source frame spectra of GRBs listed in Table \ref{tab1}. The
observed energy band is shown as the solid line in each case.
$Q_{pz}$ values are marked;
solid dots for long GRBs, solid stars for short GRBs and solid triangles for
XRFs.}
\label{fig1}
\end{figure}

\begin{figure}[!htp]
\begin{center}
\includegraphics[height=15cm,angle=-90]{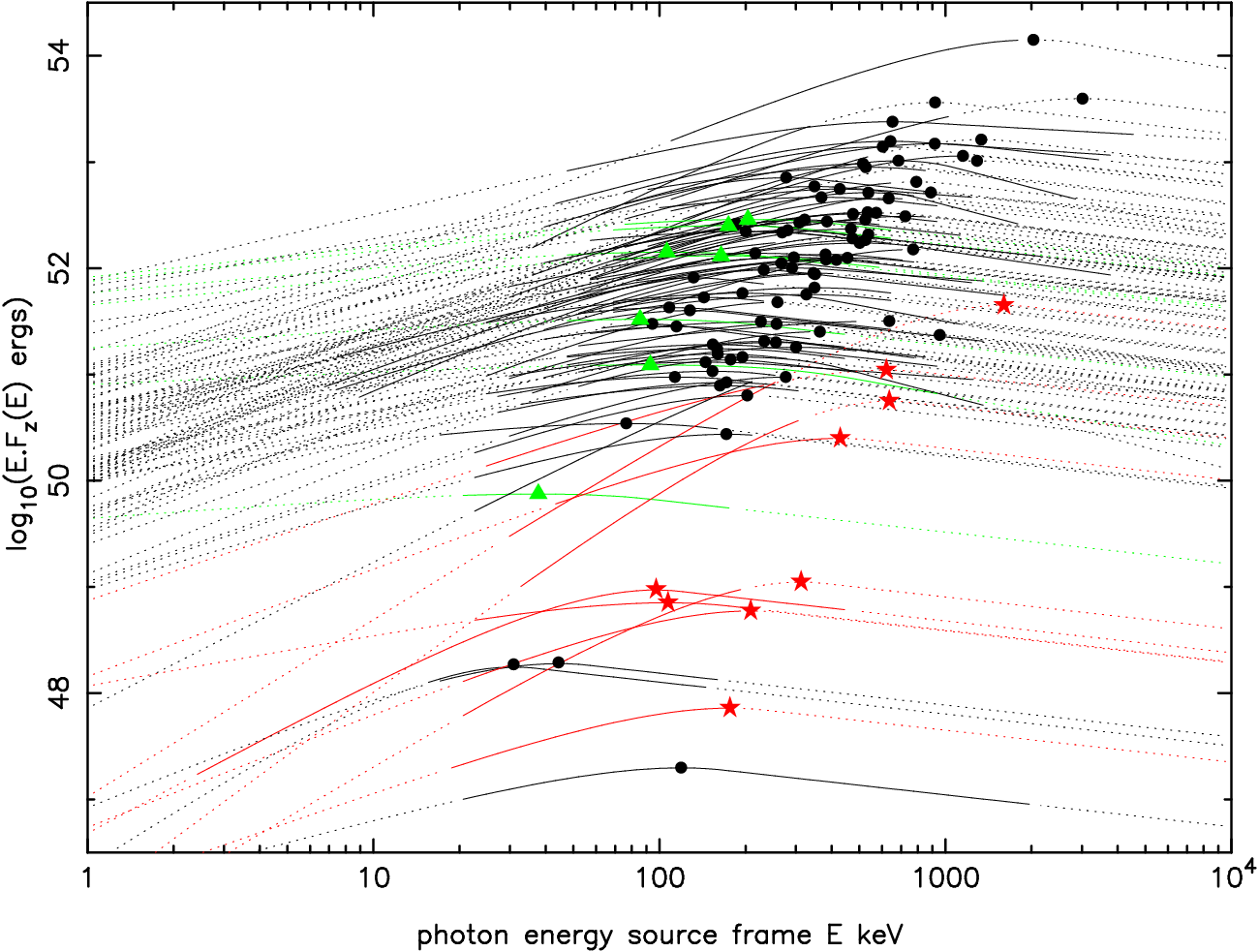}
\end{center}
\caption{The 1 keV to 10 MeV
$E.F_{z}(E)$ source frame spectra of GRBs listed in Table
\ref{tab1}. The
observed band is shown as the solid line in each case.
$E_{pz}Q_{pz}$ values are marked at the peak energy $E_{pz}$;
solid dots for long GRBs, solid stars for short GRBs and solid triangles for
XRFs.}
\label{fig2}
\end{figure}

Figure \ref{fig4} shows the Amati relationship
for the GRBs in Table \ref{tab1}.
Here and subsequently we use the exact form for $E_{iso}$, calculated
from $I_{bol}$, and not the
approximation involving $I_{fit}$ which was only introduced to derive
Equation \ref{eq11}.
The correlation line shown (derived
ignoring the obvious outliers) is $E_{pz}\propto E_{iso}^{0.46}$ 
consistent with Amati 2006, $E_{pz}\propto E_{iso}^{0.5}$.
All the short bursts are outliers with
low $E_{iso}$ values compared with the long bursts of similar
$E_{pz}$ value. The other notable outliers are GRB980425
and GRB060218 (see Amati 2006, Campana et al. 2006).
The XRFs (characterised by the hardness ratio of the low energy spectra, see
below) all fall on the lower edge of the correlation with low $E_{pz}$ 
compared with $E_{iso}$.
A more fundamental
relationship is that between the flux density $Q_{pz}$ and the
characteristic energy $E_{wz}$ which is also shown in Figure \ref{fig4}.
It appears that, disregarding the short bursts, the Amati correlation is
tighter than this new relationship but this is deceptive.
Unlike $E_{iso}$ and $E_{pz}$, $Q_{pz}$ and $E_{wz}$
are independent and their product provides the isotropic
energy $E_{iso}$ (Equation \ref{eq7}). We now have a correlation which
goes beyond the simple fact that GRB spectra have the Band function profile.
Crudely, $Q_{pz}$ is a measure of the
height of the spectrum as plotted in Figure \ref{fig1} and $E_{wz}$
(which is itself a function of $E_{pz}$, $\beta_{X}$ and $\beta_{\gamma}$)
is a measure of the characteristic photon energy.
There is a weak correlation between these two
quantities, $E_{wz}\propto Q_{pz}^{0.3}$, as can be seen
in Figures \ref{fig4} and \ref{fig1}.  However,
the pattern of outliers is the same as for the Amati relationship.
The short bursts, and the sub-luminous long burst, GRB980425,
have significantly low $Q_{pz}$ values but $E_{wz}$ values
which are comparable to the gamut of long bursts. 
The bursts designated as XRFs (see below)
all lie in the low tail of the $E_{wz}$ range
but have $Q_{pz}$ values which are similar to many long bursts.

\section{The rate profile and luminosity time of the prompt emission}

The analysis above has highlighted the well known problems associated with
the Amati relation and other correlations involving $E_{iso}$.
We now consider a way of converting $E_{iso}$ into a characteristic
luminosity to see if this can improve the situation.
The variety of time variablity in the prompt emission from GRBs is 
astonishing. Some bursts consist of a single Fast Rise Exponential Decay
(FRED) profile, other have multiple peaks, some are very spikey with
rapid variations while others have a smoother profile. The luminosity
is continually varying between bright, short peaks and low troughs
and in some cases the flux drops below the detection threshold for a while
before flaring up again. With such a range of behaviour defining some
characteristic luminosity and/or duration
is tricky. Reichart et al. (2001) showed that the peak
luminosity correlated with a variability measure $V$ 
computed by taking the difference between the light
curve and a smoothed version of the light curve
where the smoothing or correlation time was the 
time taken to emit the brightest fraction $f$ of the flux,
$T^{E}_{f}$. They showed that the most robust correlation was obtained for
$f\approx0.45$. The correlation of the peak luminosity with $T_{45}$
has been adopted by subsequent authors, for example Guidorzi et al. (2005),
Firmani et al. (2006), but in all cases
the peak luminosity must be defined using some small arbitrary bin size
(typically 1 second) and the only connection between the total fluence
and the peak luminosity is indirect, through the $T_{45}$ value.

The variability measure $V$ depends on the correlation of structures
(peaks, troughs etc.) in the light curves. Here we try a different
approach in which the sequence of features or events in the light curves
is abandoned completely. We identify the time periods in which 
significant flux is measured and then construct a {\em rate profile}
by sorting the sequence of count rate samples from these time periods
into descending order to produce,
for every GRB, a monotonically decreasing function, $f_{s}(t_{s})$, where
$t_{s}$ is sorted time. The total sum of
all the samples should be the total count fluence and the profile
is normalised by dividing by this fluence so that the integral under
the profile is unity. Such a rate profile shows what fraction of the
burst is spent at what fraction of the peak rate and has the general
form shown schematically in Figure \ref{fig5}.
Examples of these
rate profiles are shown in Figures \ref{fig6} and \ref{fig8}.
The time periods in which significant flux is detected were found by
successive correlation with boxcar functions of increasing width. It
doesn't matter if the total duration of these
periods is a little larger than required to capture the total fluence
because the small excess of samples in the tail can be dropped and
the rest of the profile is unchanged.
Remarkably the shape of these rate profiles is surprisingly similar
for {\em all} GRBs and is insensitive
to the time bin size used as long as it is not too large or
too small. If the bin size is too large
then there may be too few samples defining the profile, but we found that
a number of bins $>20$ was fine. Using excessively large time bins can
also hide significant real structure in the fluctuations of the light curve
and this should be avoided.
At the other extreme, if the bins are
too small the number of counts per bin may drop to single figures and
the profile shape is again compromised. In practice all long bursts are
well represented using $\approx64$ ms bins while short bursts require 
$\approx4$ ms bins or something similar.

The influence of statistical fluctuations (noise) on the rate
profiles is rather strange. Because the integral is normalised to unity
statistical fluctuations on the total fluence are not included.
The profile reflects the distribution of the detected flux over a
range of brightness but is not influenced by uncertainties in the total flux.
The sorting of bins into decreasing brightness order also ensures the profiles
are always smooth with the larger errors or distortion due to noise
accumulating at the start and end of the profile. This is often most noticable
as a slight increase in gradient or curl over at the end of the profile.
Although errors can be estimated for each of the samples, $d_{i}$,
Chi-squared minimization using
these errors cannot be employed for any function fitting because
the sorting operation destroys the meaning of the errors. i.e. the scatter of
the sorted data values about the fitted function is not governed directly by
the errors on $d_{i}$.

Most profiles are well represented by an empirical function of the form
\begin{equation}
f_{s}(t_{s})=f_{0}\left(
1-\left( \frac{t_{s}}{T_{E}}\right)^{1/C_{L}}\right)^{C_{L}}+f_{E}
\label{eq12}
\end{equation}
where $T_{E}$ is the total emission time or duration of the profile,
$f_{E}$ is the level of the profile at $T_{E}$ and represents the
minimum detectable flux (or luminosity) and
$C_{L}$ is a luminosity index which describes the curvature.
This function is illustrated in Figure \ref{fig5}.
Because the profile integral
is normalised to unity the peak value at the start is
$f_{0}+f_{E}=1/T_{L}$, where
$T_{L}$ is a {\em luminosity time} in seconds. The peak flux is then given by
the fluence divided by the luminosity time, $F_{tot}/T_{L}$ cts s$^{-1}$
or, perhaps more intuitively, the peak flux multiplied by $T_{L}$
is the total fluence. Because $T_{L}$ is derived from the functional
fit of all the data it does not depend strongly
on the time bin size (as discussed
above) and therefore the peak flux calculated using $T_{L}$ is also not
dependent on the binning.
If $C_{L}=1$ the
profile is linear and if $C_{L}>1$ the profile is concave and the fraction
at high rate is smaller.
If $C_{L}<1$ the curvature would be negative
but this is not seen for any GRBs. So $C_{L}$ is a measure
of the sharpness or spikiness of the profile.

We fitted all rate profiles with the function $f_{s}$ given
by Equation \ref{eq12} finding the best fit values for the parameters
$T_{L}$, $C_{L}$ and $f_{E}$ using a least squares statistic
\begin{equation}
\Sigma=100N\Delta t^{2} \sum(d_{i}-f_{si})^{2}
\label{eq13}
\end{equation}
where $N$ is the total number of samples, $d_{i}$, of time width $\Delta t$.
Note $T_{E}$ is fixed as the cumulative duration of all the significant
samples detected, $T_{E}=N\Delta t$.
The $\Sigma$ statistic has properties similar to
reduced Chi-Squared, with
typical values in the range 0.5-2.0 (set by the scaling factor of 100),
independent of the number of samples $N$ or the sample size $\Delta t$.
Table \ref{tab2} provides a 
complete list of all the temporal parameters. This table also includes
the instrument and a GRB classification using the usual observational
definitions:
Short bursts if $T_{90}<2$ s and X-ray Flashes (XRFs) if
fluence(1-30 keV)/fluence(30-500 keV)$>1$.

Figure \ref{fig6} shows examples of typical fits. Note that
sorted time is scaled by $1/T_{E}$ and the $f_{s}$ values by $T_{L}$ so
that both axes take the range 0-1. The top-right panels show
GRB021211 which is a typical FRED burst and has a low curvature index,
$C_{L}=1.28$. The top-left and bottom right panels show
GRB990510 and GRB070521 which have more complicated flaring structure but
are well fitted with $C_{L}$ values of 2.48 and 1.69 respectively.
The remaining objects have short bright spikes and extended low
level emission, a class discussed by
Norris \& Bonnell (2006). GRB050724 and GRB051221A are essentially short bursts
followed by a low level, extended tail and the combination of these
features produces large $C_{L}$ values, 3.17 and 2.85
respectively.
For GRB051221A, $\Sigma=3.22$ which is rather high. In this
case the short spike followed by the extended tail produces an extra feature
or wiggle in the rate profile which is not fitted by the
simple $f_{s}$ function, Equation \ref{eq12}.
For these and similar bursts a sample size of 4 ms was used to
accomodate the profile of the initial short spike.
The left-hand panel of
Figure \ref{fig7} shows the distribution of $\Sigma$ and $T_{L}$
values for all GRBs in Table \ref{tab2}. There is no correlation between
the goodness of fit measured by $\Sigma$ and the luminosity time, $T_{L}$.
The same is true for $\Sigma$ and the luminosity index $C_{L}$.
Figure \ref{fig8} shows the worst fits
of rate profiles with large $\Sigma$ values. In all these GRBs
the peak value, $1/T_{L}$, is a good approximation to the data peak
but the fit is compromised by undulating features. GRB990705 and
GRB061007 represent a small group of bursts which have flares that
rise fast, are reasonably flat at the top and decay fast. These produce
a characteristic S-feature in the profile.
Only 12 rate profiles (out of 101) have $\Sigma>2$ and
only 3 of these have a substantial mis-match, GRB990705, GRB061007 and
GRB061210. The latter is an extreme example of a short burst, $T_{L}=0.03$ s,
which has an extended low flux tail giving $T_{90}=85.3$ s. We note that
GRB991216 has a faint pre-cursor just visible on the lightcurve plot.

The combination of luminosity time, $T_{L}$, and curvature index,
$C_{L}$, gives us information closely related to $T_{45}$. The right-hand
panel of Figure \ref{fig7} shows the correlation between $C_{L}$ and the
ratio of $T_{45}$ calculated directly from the sample values $d_{i}$ and
$T_{L}$ from the fitted function. $T_{45}$ could be calculated
by integration of the fitted function using the parameters
$T_{L}$, $C_{L}$, $f_{E}$ and $T_{E}$ and this would produce a smooth curve of
$C_{L}$ vs. $T_{45}/T_{L}$ if $f_{E}$ were zero or constant.
The parameter $C_{L}$, for example,
could be replaced by $T_{45}$ and the fitted function would still be
uniquely defined. The scatter in Figure \ref{fig7}
results from the small differences between the data and the fitted function
and the value of $f_{E}$ which is generally much smaller than
$1/T_{L}$ but different for each GRB.
Error ranges for $T_{L}$ and $C_{L}$ were estimated assuming
the statistic $\Sigma$ has properties similar to reduced Chi-Squared.
The errors so derived are not statistically correct, because of
the odd statistical nature of the sorted rate profile, and in some
cases they are an over estimate as is evident from the
scatter in Figure \ref{fig7}.

Although the minimum flux level, $f_{E}$, was included in the fitting it
is a measure of the instrument sensitivity rather than some intrinsic
property of the rate profile. If the noise level were lower the
number of significant samples detected would increase, $T_{E}$ would
get bigger and $f_{E}$ would decrease. The instrument would detect
a slightly larger fluence, $F_{tot}$, and the fitted value of
$T_{L}$ would increase a little, however, the peak flux level,
$F_{tot}/T_{L}$ would remain unchanged and $C_{L}$ would be
essentially the same. The analysis of the rate profile
described above provides a robust estimate of the peak flux (or peak
luminosity) using all the available light curve data and is
not biased by the instrument sensitivity providing the burst
detection significance is secure in the first instance.
The error on the peak flux so estimated is dominated by the error on the
fluence rather than any error associated with estimating the
luminosity time, $T_{L}$. It is also unchanged by the choice of sample
size, $\Delta t$, providing the number of samples is sufficient to
capture the details of the emission profile as already discussed above.
We can never be sure that resampling a light curve with a smaller
$\Delta t$ will not reveal a very short, bright, isolated spike which was
hidden by the previous binning and this would compromise the shape
of the profile, but such has not been seen in any of the GRB light
curves analysed so far (about 250 including all {\em Swift} bursts to date).

Using the redshift, $z$,
we can calculate the luminosity time in the source frame,
$T_{Lz}=T_{L}/(1+z)$ and $90\%$
duration in the source frame, $T_{90z}=T_{90}/(1+z)$.
The peak luminosity multiplied by the $T_{Lz}$ gives the
isotropic energy, $L_{iso}T_{Lz}=E_{iso}$ ergs. This simple
property of $T_{Lz}$ makes it a highly significant measure of the burst
duration and is why we chose to call it the luminosity time. Such a time
is often introduced in theoretical dicussions, see for example $t_{j}$
in Thompson et al. (2007) or $t_{burst}$ in Ghirlanda et al. (2007).
Above we have described a method to calculate this time for every GRB.

\section{Characterisation of the prompt emission in the source frame}

The prompt emission of each GRB in the source frame is characterised
by the peak energy density,
$Q_{pz}$ ergs keV$^{-1}$, the characteristic photon energy, $E_{wz}$ keV
(which embodies the spectral indices $\beta_{X}$, $\beta_{\gamma}$ and the
peak energy $E_{pz}$, Equation \ref{eq8}),
the luminosity time, $T_{Lz}$ s, and the luminosity curvature, $C_{L}$.
Figure \ref{fig9} shows $T_{Lz}$ plotted against the standard measure of
burst length $T_{90z}$ where the dashed line shows equality.
For bursts consisting
of a single smooth pulse then $T_{Lz}\approx T_{90z}$. If there is more
structure in the light curve and, in particular, if there are periods when
the flux drops to zero then $T_{Lz}<T_{90z}$. In some cases a short
precursor pulse is followed by a long time gap before the main burst
starts and then $T_{Lz}<<T_{90z}$. So the ratio of the two times is a
crude measure of the variability but this
includes all time scales and long periods when no flux is detected
and is not equivalent to the short time scale variability defined by
Reichart et al. (2001).
The top-right panel
of Figure \ref{fig9} shows the distribution of $T_{Lz}$. Two
peaks containing the short-bursts,
centred around 0.05 seconds, and long-bursts centred
at 5 seconds, are clearly visible.
The distribution of $C_{L}$ is shown in the lower left-hand panel of
Figure \ref{fig9}. Most bursts are contained in a symmetrical peak
centred on $C_{L}=1.6$. The few bursts with $C_{L}>2.2$ include
the short bursts which have a long weak tail and bursts which exhibit
several very short spikes on top of a more generally smooth emission.
The bottom right-hand panel of Figure \ref{fig9} shows the distribution
of the Band lower energy spectral index,
$\beta_{X}$. Hard bursts have $\beta_{X}<0$ and softer bursts have
$\beta_{X}>0$. We do not show the distribution of the high energy
spectral index, $\beta_{\gamma}$,
because this parameter is only available for a few bursts and in most
cases it was set to $\beta_{\gamma}=1.3$ which is the approximate average
found by {\em BATSE}.

The distributions of the remaining parameters,
the characteristic photon energy, $E_{wz}$, and the peak energy density,
$Q_{pz}$, are shown at the top of the Figure \ref{fig10}.
$E_{wz}$ stretches over two decades from 100 keV to 10000 keV.
$Q_{pz}$ has a much larger spread with a main peak spanning three
decades and a low energy tail covering another three.
Since the product of the two gives us $E_{iso}$ the
range of isotropic energy is very large, as is evident from Figure \ref{fig3}
and the Amati relation plotted in Figure \ref{fig4}.
The peak energy density, $Q_{pz}$, is correlated with the luminosity
time $T_{Lz}$ as demonstrated by the bottom left-hand panel of Figure
\ref{fig10}.
Short bursts have $Q_{pz}<10^{48.5}$ while in general
long bursts have larger $Q_{pz}$ values. The two notable exceptions
are, as before, GRB980425 and GRB060218 which are long bursts with very
low luminosity. Five short bursts with long tails that are classified
as long because their $T_{90}>2$,
GRB050603, GRB050724, GRB061006, GRB061210 and GRB070714B
have $T_{Lz}$ of 0.22, 0.38, 0.33, 0.02 and 0.48 s
respectively and these sit below the main long grouping along with the
shorts.
The XRFs tend to have lower
$Q_{pz}$ and lower $T_{Lz}$ values within the long burst population.
The peak luminosity density of a burst
is given by $Q_{pz}/T_{Lz}$ ergs keV$^{-1}$ s$^{-1}$. This has
a much narrower distribution than either $Q_{pz}$ or $T_{Lz}$ with
a $90\%$ range just over 2 decades, $3.1\times10^{47}-7.8\times10^{49}$
ergs keV$^{-1}$ s$^{-1}$, as is
clear from the histogram in the bottom right-hand panel of Figure \ref{fig10}.
Both short and long bursts have similar values of peak luminosity
density (the short bursts are shown as the white histogram, all bursts
are shown in the grey histogram).

We have used Principal Components Analysis (PCA) to investigate the
scatter within the parameter space described above
($T_{Lz}$, $C_{L}$, $Q_{pz}$ and $E_{wz}$). This analysis confirms that
there is indeed a correlation between $T_{Lz}$ and $Q_{pz}$ and the
best fit is very close to proportionality with index $0.89$ but there is
considerable scatter with Pearson's correlation coefficient is $r=0.5$,
Kendall's $\tau=0.3$, $4.5\sigma$ (see Figure \ref{fig10}).
If we fix this index to unity then the only other significant
correlation is between $Q_{pz}/T_{Lz}$ and the characteristic energy
$E_{wz}$.
This is shown in the left-hand panel of Figure \ref{fig11} including
the best fit correlation,
$E_{wz}\propto (Q_{pz}/T_{Lz})^{0.25}$ which has Pearson's 
correlation coefficient $r=0.55$ and Kendall's $\tau=0.32$, significance
$4.8\sigma$.
The critical difference between this
plot and the right-hand panel of Figure \ref{fig4} (the Amati
relation) is that the 
peak energy density $Q_{pz}$ has been converted to a peak luminosity
density by dividing by the time $T_{Lz}$. The large difference between
the short and long bursts has disappeared and most bursts are now 
clustered in a small area on the energy-luminosity plane.
It seems that all correlations
involving the properties of GRBs must have outliers and this is no
exception; GRB980425 still refuses to conform but the remaining 100 bursts come
into line.

The correlation shown in Figure \ref{fig11} between the peak luminosity density
and the characteristic photon energy in the source frame
is the first GRB relationship to unify the short and the long
bursts.
If the peak luminosity density is multiplied by $E_{wz}$ the x-axis
becomes the peak isotropic luminosity, $L_{iso}$ ergs s$^{-1}$.
The correlation of $E_{wz}$ vs. $L_{iso}$ is shown in the
right-hand panel of Figure \ref{fig11} along with the best fit
\begin{equation}
\frac{E_{wz}}{381\:{\rm keV}}
=\left(\frac{L_{iso}}{10^{50}\:{\rm ergs\:s^{-1}}}\right)^{0.25}
\label{eq14}
\end{equation}
which has a Pearson's correlation coefficient of $r=0.75$, Kendall's 
$\tau=0.54$, significance $8.0\sigma$. 
Thus Figure \ref{fig11} encapsulates a major result of this work, showing
a high quality correlation of characteristic photon energy with
peak isotropic luminosity for 101 GRBs including 9 short bursts and 7 XRFs.
The correlation between $E_{wz}$ and $L_{iso}$ is similar to those
reported by Yonetoku et al. (2004) and
Firmani et al. (2006) but there are important differences. Here we have
estimated the peak isotropic
luminosity from the rate profile so we are not
restricted to long bursts or a particular time bin size, and the
peak energy, $E_{pz}$, is replaced by the characteristic photon energy,
$E_{wz}$. We note that the correlation derived by Yonetoku et al. (2004)
is significantly steeper, $E_{pz}\propto L_{iso}^{0.5\pm0.1}$
but they used a rather small sample of 16 GRBs.
We can identify 13 of these objects
in our sample and we find they give $E_{pz}\propto L_{iso}^{0.41\pm0.06}$ with
Pearson's correlation coefficient $r=0.81$, consistent with their result.
The same 13 objects also give 
$E_{wz}\propto L_{iso}^{0.31\pm0.04}$ with $r=0.88$
so using $E_{wz}$ in place of $E_{pz}$ gives a slightly tighter correlation 
with a shallower slope which is consistent with our result obtained
from the full sample of 101 bursts.
Unlike the Firmani et al. relationship the present
correlation does not contain $T_{45z}$. We tried including $T_{45z}$ in the
PCA but found no significant correlation or reduction in the scatter.
If we replace $E_{wz}$ by $E_{pz}$ in the PCA of the complete sample then
$E_{pz}\propto L_{iso}^{0.27}$ with $r=0.71$ and Kendall's $\tau=0.52$,
significance $7.7\sigma$ so, again, using $E_{wz}$ yields a tighter
correlation with a shallower slope compared to $E_{pz}$. The small change in
slope arises because the correlation of $E_{pz}$ with $E_{wz}$ is not quite
unity (see Figure \ref{fig3}).

For each burst we calculate $K_{z}$ which is a measure of its displacement
perpendicular from the the best fit correlation line in the right-hand panel
of Figure \ref{fig11}.
\[K_{z}
=\left(\frac{E_{wz}}{1320\:{\rm keV}}\right)^{0.97}
\left(\frac{L_{iso}}{1.45\times10^{52}\:{\rm ergs}\:{\rm s}^{-1}}\right)^{-0.24}\]
\begin{equation}
=\left(\frac{E_{wz}}{1320\:{\rm keV}}\right)^{0.74}
\left(\frac{Q_{pz}}{2.08\times10^{49}\:{\rm ergs}\:{\rm keV}^{-1}}\:
\frac{1.89\: {\rm s}}{T_{Lz}}\right)^{-0.24}
\label{eq15}
\end{equation}
This is a function of the ratio of the characteristic photon energy to the peak
isotropic luminosity.
The constants quoted in this definition are the mean values of the parameters
so they represent the centre of the clustering of objects within the
parameter space.
The distribution of $K_{z}$ is plotted in Figure \ref{fig12}.
The mean value is $\log_{10}(K_{z})=0$ or, equivalently,
$K_{z}=1$.
Hard-dim bursts (including most short bursts)
have $K_{z}>1$, soft-bright bursts (including all XRFs)
have $K_{z}<1$. The distribution is approximately log-normal
(the best fit Gaussian profile is shown in Figure \ref{fig12})
and has a rms width of
$\sigma[\log_{10}(K_{z})]=0.19$. $90\%$ of the GRBs (90 objects) are
contained in the range $0.45<K_{z}<1.95$.
The obvious outlier is GRB980425/SN1998bw which is either very 
sub-luminous or has an exceptionally high peak energy for such a dim
burst.
Under the hypothesis
that $K_{z}$ is constant, $\chi^{2}=421$ with 99 degrees of freedom
and the mean of the estimated errors on $\log{10}(K_{z})$ is 0.12
so there is clear evidence for intrinsic scatter in $K_{z}$ with an
estimated $90\%$ range of $0.57<K_{z}<1.75$.
The largest uncertainties arise from the estimation of $E_{wz}$ because
this depends on $E_{pz}$ and the spectral indices $\beta_{X}$, $\beta_{\gamma}$.
The mean value of $K_{z}$ for the pre-{\em Swift} bursts is
$-0.05$ and for {\em Swift} bursts is $0.02$ so they are statistically
indistinquishable.
The distribution for pre-{\em Swift} bursts, plotted as the white 
histogram in Figure \ref{fig12}, sits symmetrically within the total 
distribution.
The right-hand panel of Figure \ref{fig12} shows $\log_{10}(K_{z})$
as a function of redshift, $z$. There is no obvious trend. The objects
with the smallest errors that contribute most to the high $\chi^{2}$ show
no dependence on redshift. Table \ref{tab3} provides a complete listing
of the rest frame parameters and associated errors.

\section{Discussion}

Within the new parameterisation of the temporal and spectral properties
of the prompt GRB emission the three important quantities are the
characteristic energy in the source frame, $E_{wz}$ keV (Equation \ref{eq8}),
the energy density at the peak of the $E.F_{z}(E)$ spectrum,
$Q_{pz}$ ergs kev$^{-1}$ (derived from  the total fluence, Equations \ref{eq2},
\ref{eq5} and \ref{eq6}) and the luminosity
time, $T_{Lz}$ s, derived from the rate profile.
The ratio $Q_{pz}/T_{Lz}$ gives us the peak luminosity density in
ergs keV$^{-1}$ s$^{-1}$ where ``peak'' corresponds to both the
maximum in the $E.F_{z}(E)$ spectrum and the maximum flux level in
the light curve.
$E_{wz}$ and $Q_{pz}/T_{Lz}$ are correlated and it is this correlation
which gives rise to the Amati relation (Amati et al. 2002).
The instantaneous maximum brightness
of the prompt emission is characterised by a function of the
photon energy/peak luminosity ratio $K_{z}$ given
in Equation \ref{eq15}. This is not a constant but it covers a remarkably
small dynamic range compared with the constituent parameters, $E_{wz}$,
$Q_{pz}$ and $T_{Lz}$. Given the measurement errors it is difficult to 
make an accurate estimate of the intrinsic dynamic range but it is 
certainly less than $0.5<K_{z}<2.0$ and this holds for 100 GRBs in the
sample of 101 we have analysed including long, short and XRFs, the
exception being GRB980425.

\subsection{Is $K_{z}$ intrinsic?}

We might wonder whether the narrow range in $K_{z}$ is an artefact of the
observational data or something instinsic to the nature of the GRB emission?
An artificial tightness of the energy-luminosity correlation could arise
in several ways; the observed quantities may be correlated by some property 
of the instrumentation/measurement,
the measured positions of GRBs in the energy-luminosity
plane could be incorrect because of some systematic error/bias
or GRBs from certain areas in the plane
may be selectively missed. The measured quantities in the observer frame which
map to $E_{wz}$ and $Q_{pz}/T_{Lz}$ are the peak energy, $E_{p}$ keV, and
the spectral energy density at the peak
\begin{equation}
f_{p}=\frac{N_{tot}}{T_{L}}\exp[(1-\beta_{X})(E_{p}^{-1}-1)] E_{p}^{-\beta_{X}}
\label{eq16}
\end{equation}
ergs cm$^{-2}$ keV$^{-1}$ s$^{-1}$. These are plotted in the top left-hand
panel of Figure \ref{fig13}. There is no tight clustering or significant
correlation. Pearson's correlation coefficient is $r=0.27$
and Kendall's $\tau=0.13$, significance $2.0\sigma$.
The $90\%$ range of $f_{p}$ is $4.1\times10^{-10}-2.8\times10^{-8}$
ergs cm$^{-2}$ keV$^{-1}$ s$^{-1}$ 
and the $90\%$ range of the observed peak energy is
$34-412$ keV, both one to two orders of magnitude. Using the
redshift to transform these into $Q_{pz}/T_{L}$ and $E_{wz}$ produces the
distributions shown in Figure \ref{fig10}. These quantities have slightly
narrower distributions in the source frame with $90\%$ ranges of
$3.1\times10^{47}-7.8\times10^{49}$ ergs keV$^{-1}$ s$^{-1}$ and
$511-3450$ keV respectively.
Finally they combine in $K_{z}$ which has a rather narrow
$90\%$ range of $0.45-1.95$ and some of this is attributable to the
measurement errors. It is very unlikely that some systematic error or bias
in the measured quantities which have a large dynamic range and are not
correlated
conspires to give such a tight correlation and we conclude that $K_{z}$
encodes real, useful, information about the source frame properties of
the prompt emission. The rate profile fitting not only provides us with 
the peak flux density level but also 
the minumum detected flux density
\begin{equation}
f_{m}=N_{tot}f_{E}\exp[(1-\beta_{X})(E_{p}^{-1}-1)] E_{p}^{-\beta_{X}}
\label{eq17}
\end{equation}
ergs cm$^{-2}$ keV$^{-1}$ s$^{-1}$. The top
right-hand panel of Figure \ref{fig13}
shows $f_{m}$ vs. $f_{p}$. There is some clustering of the weaker bursts
along the line $f_{p}\approx5f_{m}$ and clearly the area above this
line in the top left corner is below the threshold.
It could be that we are preferentially missing hard-dim bursts
while soft-dim bursts are detected
but redshift works in our favour because the distant dim bursts
are redshifted into the lower observation energy band where the
sensitivity is higher and time dilation stretches the light curve so
we have longer to detect the emission. We are undoubtedly
missing low luminosity bursts especially at high redshift.
The lower panels of Figure \ref{fig13} shows $L_{iso}$ and $E_{wz}$
plotted vs. redshift.
The hard (high $E_{wz}$) and most luminous
(high $L_{iso}$) sources are seen at all redshifts
while the softer, weaker sources are only seen at low redshifts,
entirely as one would expect,
but as we can see from Figure \ref{fig12} there is no obvious
difference in the distribution of $K_{z}$ as a function of $z$. There is
no reason to suspect that absence of dim bursts too weak to detect
is biasing the distribution in the photon energy-luminosity plane.

Our conclusions are somewhat different from Butler et al. (2007).
The combination of recent {\em Swift} detections with pre-Swift results
confirms the general correlation between $E_{pz}$ and $E_{iso}$ (the Amati
relation) but the spread is indeed large and many bursts, including 
all the short bursts, are extreme outliers from the bulk correlation of
the long bursts. Such a correlation is, in part, a simple consequence of
the shape of a typical GRB spectrum (the Band function)
but the spread and presence of many outliers renders this correlation 
insensitive in testing of cosmological world models.
There is a clustering of events
in the ratio of fluence/duration (effectively luminosity), however, we think 
the root of the problem is not observational bias
or sensitivity thresholding but rather that the Amati relation
(and similar correlations involving $E_{iso}$) are looking at the
wrong parameter space. Yes, this is the demise of the existing pre-{\em Swift}
high-energy correlations, but if we re-cast them in terms of the
instantaneous peak luminosity and we replace $E_{pz}$ by $E_{wz}$, which
combines the spectral parameters $E_{pz}$, $\beta_{X}$ and $\beta_{\gamma}$,
then they reappear in a new light. The quantities in the observer frame
are not correlated, the short and dim bursts are no longer
outliers in the source frame
correlation and there is no difference between {\em Swift}
and pre-{\em Swift} detections.

\subsection{Correlations involving evolution of parameters}

All the analysis presented here involves average spectral and temporal
properties of the prompt emission.
$E_{wz}$ and $Q_{pz}$ are derived from the time-integrated
spectra and $T_{Lz}$ is estimated from the full energy band light curves.
We know that GRB spectra evolve with time and the light curves are different
in different energy bands. In general the spectra soften as the burst proceeds,
$\beta_{X}$ increases and $E_{p}$ decreases with time (e.g. Goad et al. 2007,
Page et al. 2007). The light curves
are shorter and more spikey at high energies than they are at low energies
(Reichart et al. 2001).
The lag-luminosity (Norris et al. 2000) and variability-luminosity
(Reichart et al. 2001) correlations are testament to this temporal-spectral
evolution. In this work
we have estimated the hardness and brightness of just the peak
emission. If instrumentation could follow the evolution of the
characteristic energy
and luminosity through the light curve each burst would form a track on the
energy-luminosity density plane which may run from top right to bottom
left with $K_{z}\sim constant$.
The lag and variability correlations may provide a means by which
scatter can be introduced
in $K_{z}$ although physical reasons for this are not immediately
apparent. Further analysis and better quality data are
required to explore the evolution of $K_{z}$ through individual bursts.

\subsection{Emission processes}

The correlation between the hardness and brightness of GRB spectra, 
previously in the form of the Amati relation and now the correlation
between $E_{wz}$ and $Q_{pz}/T_{Lz}$, is a challenge to theoretical
modelling of the prompt emission. Within the standard fireball picture
there are many variants involving internal and external shocks in which
synchrotron emission, inverse Compton scattering and photospheric
emission feature, and the fireball itself may be dominated by kinetic energy
or magnetic energy (Poynting flux). The initial problem is to predict
a spectrum which has the general form of the Band function with a spectral
break or curvature characterised by some energy, $E_{c}$, $E_{p}$ or $E_{wz}$,
and the second problem is to
predict the coupling between the characteristic energy or hardness of the
spectrum and the luminosity (see the review by Zhang \& M\'{e}sz\'{a}ros 2002).

With a kinetic energy dominated
outflow and a simple synchrotron model generated by internal shocks,
incorporating a peak in the electron energy
one expects $E_{pz}\propto \Gamma^{-2}t_{var}^{-1}L^{1/2}$ where
$\Gamma$ is the bulk Lorentz factor, $t_{var}$ is the typical variability
time scale associated with the internal shocks
and $L$ is the luminosity (Zhang \& M\'{e}sz\'{a}ros 2002).
This is consistent with the Amati
relation if $L\propto E_{iso}$ (which is not the case if we include
both short and long bursts as shown above)
and there is a constancy of both $\Gamma$ and $t_{var}$
across all bursts, which seems unlikely (Rees \& M\'{e}sz\'{a}ros 2005). 
The relationship $E_{wz}\propto L_{iso}^{0.25}$
derived here and shown to hold for all bursts
is significantly flatter than the Amati relation.
If the Lorentz factor depends on luminosity, $\Gamma\propto L^{\beta}$, then
we can choose $\beta\approx1/8$ to match the observed correlation
providing $t_{var}$ is independent of luminosity and approximately
constant for all bursts. It is not obvious why the Lorentz factor should
have such a specific and low dependence on luminosity and, again,
why $t_{var}$ should be constant when the burst durations ($T_{90z}$ or
$T_{Lz}$) have such a large dynamic range. Within this model the radius
at which the emission occurs is given by $r\sim ct_{var}\Gamma^{2}$ so if
we assume typical values of
$t_{var}\sim0.01$ s and $\Gamma\sim300$, $r\sim3\times10^{13}$ cm.
Furthermore,
by considering the onset of X-ray afterglows observed by {\em Swift}
Kumar et al. (2007) estimate that emission originates at much larger
radii, between $10^{15}$ and $10^{16}$ cm, and suggest that 
synchrotron/inverse Compton parameters cannot account for the prompt 
emission.

Alternatively, we can can consider a thermal origin for the peak in
the $E.F(E)$ spectrum and the correlation of the characteristic energy
with luminosity, see for example Rees \& M\'{e}sz\'{a}ros (2005), Ryde (2005)
and Ghirlanda et al. (2007). If the photosphere of the expanding fireball has
radius $R_{0}$, Lorentz factor $\Gamma_{0}$, a blackbody spectral component
with temperature $T_{bb}$ and isotropic luminosity fraction
$\varepsilon_{bb}$ of the total isotropic luminosity $L_{iso}$,
then the observed temperature,
$T_{obs}=(4/3)\Gamma_{0}T_{bb}$, is given by
\begin{equation}
\frac{T_{obs}}{1460\:{\rm keV}}=
\left(\frac{\Gamma_{0}\:10^{7}\:{\rm cm}}{R_{0}}\right)^{1/2}
\left(\frac{\varepsilon_{bb} L_{iso}}{10^{52}\:{\rm ergs\:s^{-1}}}\right)^{1/4}
\label{eq18}
\end{equation}
Thompson (2006). This is just the Stefan-Boltzmann law modified to account
for the relativistic expansion rate of the photosphere and it matches
the observed correlation if $E_{wz}\propto T_{obs}$ and
$\varepsilon_{bb}^{1/2}\Gamma_{0}/R_{0}$
is approximately constant for all bursts. The observed spectrum
is not a single temperature blackbody. If $\varepsilon_{bb}<<1$
then any single temperature blackbody component is diluted by non-thermal
(possibly power law or inverse Compton)
components which combine to give the Band function.
If $\varepsilon_{bb}\sim 1$ the observed spectrum must result from 
the summation of a large number of thermal components with
a spread of temperatures that give an average of $T_{obs}$,
(for example, in a manner similar to that described by Ruffini et al. 2004).
We can re-arrange Equation \ref{eq18} as
\begin{equation}
\left(\frac{\varepsilon_{bb}^{1/2}\Gamma_{0}\:10^{7}\:{\rm cm}}
{R_{0}}\right)^{1/2}=
\left(\frac{T_{obs}}{1460\:{\rm keV}}\right)
\left(\frac{L_{iso}}{10^{52}\:{\rm ergs\:s^{-1}}}\right)^{-1/4}
\label{eq19}
\end{equation}
If $T_{obs}\sim E_{wz}$ this is essentially the same as the
definition of the photon energy-
luminosity quasi-constant, $K_{z}$, and so, under this interpretation,
the scatter in the correlation
arises from variations in the fireball dimension, $R_{0}$, the Lorentz
factor, $\Gamma_{0}$, or the blackbody luminosity fraction, $\varepsilon_{bb}$.
If $K_{z}=1$ then we have an average fireball with
$R_{0}/(\varepsilon_{bb}^{1/2}\Gamma_{0})\approx10^{7}$ cm.
If $K_{z}>1$ the fireball has
a higher than average $\varepsilon_{bb}^{1/2}\Gamma_{0}$
and/or a smaller radius, $R_{0}$.
If $K_{z}<1$ the fireball has a low $\varepsilon_{bb}^{1/2}\Gamma_{0}$
and/or large
radius. With $\Gamma_{0}=300$ and $\varepsilon_{bb}=1$
then $R_{0}\approx3\times10^{9}$ cm which
is the thermalization radius (i.e. the radius of the jet or fireball
photosphere) estimated by Thompson, M\'{e}sz\'{a}ros \&
Rees (2007). This radius is much smaller than estimates arising from
the internal shock model or recent estimates involving the onset of
the X-ray afterglow (see above) so independent estimates
of the radius of the prompt emission and/or the Lorentz factor may help to
discriminate between thermal and internal shock models.
We also note that baryonic photospheres
are governed by physical argument (M\'{e}sz\'{a}ros et al. 2002)
such that the ratio $R_{0}/\Gamma_{0}$ is constrained and the relationship
between $\varepsilon_{bb}^{1/2}\Gamma_{0}/R_{0}$
and $K_{z}$ may not be as simple as we have indicated.

\subsection{$K_{z}$ as a cosmological probe}

The $E_{wz}-Q_{pz}/T_{Lz}$ correlation should be useful as a pseudo
redshift indicator 
but there is evidence for intrinsic scatter in $K_{z}$, the
errors in determining the characteristic energy and luminosity density are
large and the correlation gradient is rather low, 0.24.
In particular, reliable estimation of 
$E_{wz}$ requires an accurate measurement of the
broadband spectrum and always involves some extrapolation to cover the
source frame energy band 1-10000 keV. Without a good measurement of
the peak energy, $E_{p}$, we can't calculate a good estimate of $Q_{pz}$
from $Q_{z}$. The launch of {\em GLAST} in the near
future will hopefully provide excellent high-energy spectral measurements
which will tie down the spectral parameters more precisely.
The intrinsic scatter may arise in several ways. If the 
dominant emission mechanism is non-thermal then the coupling of 
the Lorentz factor of the expansion with the luminosity and the 
variability time associated with the internal shocks may be the root
cause. If thermal processes dominate then the ratio $R_{0}/\Gamma_{0}$ may
vary as discussed above. Because we can identify classes which
fall predominately at $K_{z}>1$ (shorts) and $K_{z}<1$ (XRFs) there is
some hope that additional parameters which distinguish these
classes may serve to narrow the distribution. In the above
discussion we made no mention of collimation or beaming of the
outflow. Since the Amati relation involving $E_{iso}$ has been transformed
into the present universal correlation involving $L_{iso}$ the simple
beaming argument that underpinned the Ghirlanda relation 
(Ghirlanda et al. 2004) is not directly applicable and currently there
is no simple physical model which links collimation to scatter
in peak luminosity or the characteristic photon energy.
However, it is not
unreasonable to suppose that collimation may introduce scatter in
the peak luminosity, and that correlation of $K_{z}$ with afterglow parameters
such as optical jet break times $t_{break}$ or the time of the
start of the final X-ray afterglow, $T_{a}$  (Willingale et al. 2007),
may be fruitful. 
In its present form $K_{z}$ is not a sensitive cosmological probe
but the signs are that it may be in the future.

\section{Conclusion}

The equivalent isotropic energy, $E_{iso}$ ergs, of a GRB can be expressed
as the product of two source frame terms, a characteristic photon energy,
$E_{wz}$ keV, calculated from
the shape of the spectrum across the range 1-10000 keV and the energy
density at the
peak of the $E.F_{z}(E)$ spectrum, $Q_{pz}$ ergs keV$^{-1}$. The correlation
trend between $E_{wz}$ and $Q_{pz}$ gives rise to the Amati relation.
By stacking the samples of a GRB light curve into
descending order we can construct a rate profile. 
The functional form of such rate profiles is
common to the vast majority of bursts.
Fitting the profile gives us a
luminosity time, $T_{Lz}$ s, a measure of the burst duration
which can be used to convert the energy
density at the peak to a luminosity density at peak, $Q_{pz}/T_{Lz}$ ergs
keV$^{-1}$ s$^{-1}$. We can calculate the peak equivalent isotropic 
luminosity as a product $L_{iso}=E_{wz}Q_{pz}/T_{Lz}=E_{iso}/T_{Lz}$
ergs s$^{-1}$.

\begin{figure}[!htp]
\begin{center}
\includegraphics[height=15cm,angle=-90]{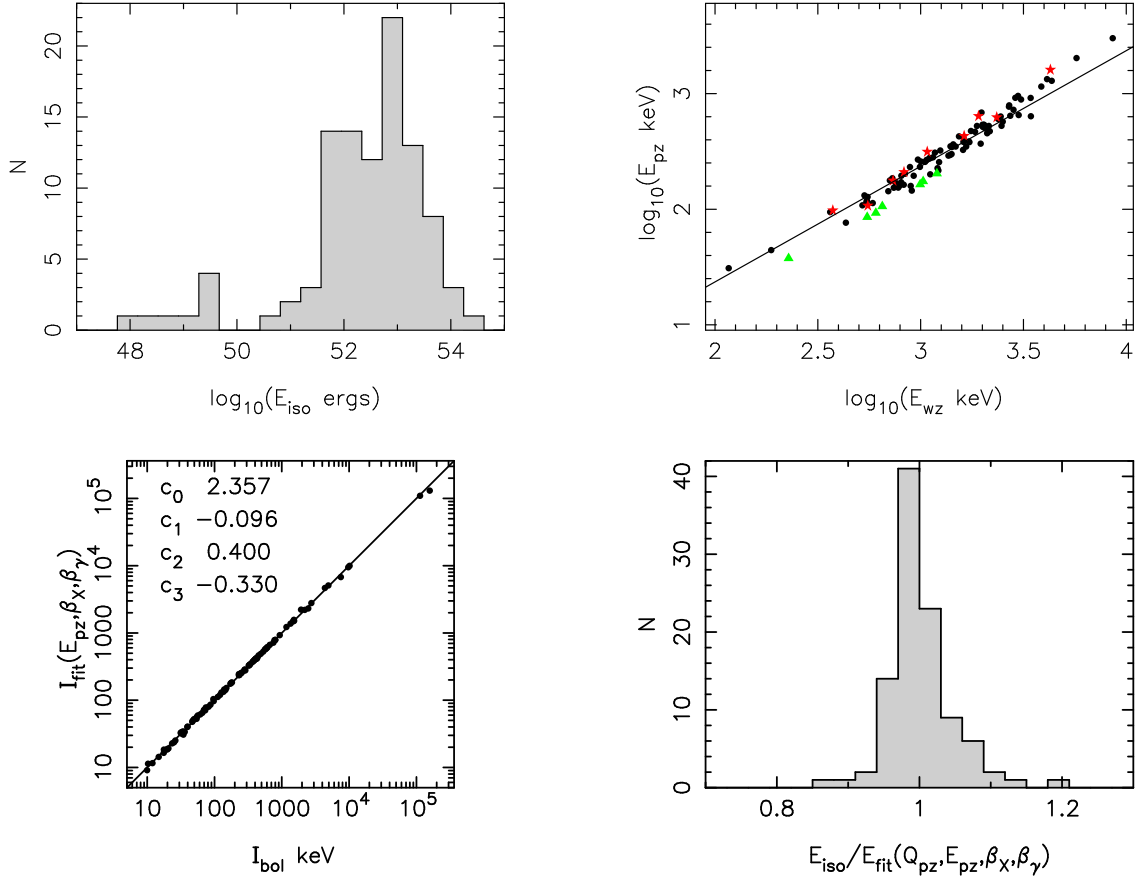}
\end{center}
\caption{Top panels: The distribution of $E_{iso}$ and the correlation
of $E_{pz}$ vs. $E_{wz}$ for GRBs listed in Table \ref{tab1};
solid dots for long GRBs, solid stars for short GRBs, solid triangles for XRFs.
The solid line is $E_{pz}=0.23 E_{wz}$ rather than the best fit correlation
which is a little steeper (see text).
Bottom panels: Comparison of the functional fit to the bolometric integral
$I_{fit}$ and the value calculated by numerical integration of the Band
function over
the interval 1-10000 keV for the GRBs in Table \ref{tab1}. The right-hand
panel shows the distribution of the
ratio of $E_{iso}$ to the value obtained using $I_{fit}$.}
\label{fig3}
\end{figure}

\begin{figure}[!htp]
\begin{center}
\includegraphics[height=15cm,angle=-90]{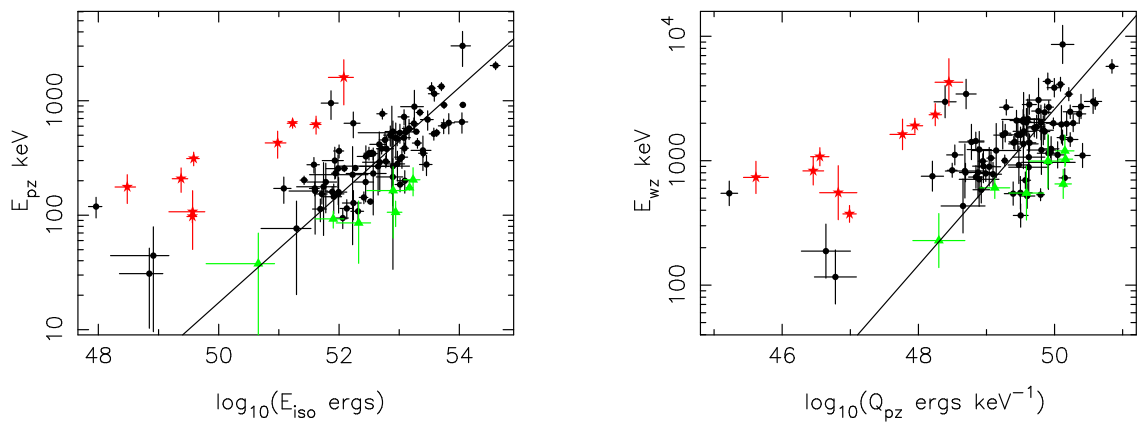}
\end{center}
\caption{Left-hand panel: The Amati relation for the GRBs in Table \ref{tab1}.
Right-hand panel: $E_{wz}$ vs. $Q_{pz}$ for the same GRBs.
Solid dots for long GRBs, solid stars for short GRBs, solid triangles for
XRFs.}
\label{fig4}
\end{figure}

\begin{figure}[!htp]
\begin{center}
\includegraphics[height=8cm,angle=-90]{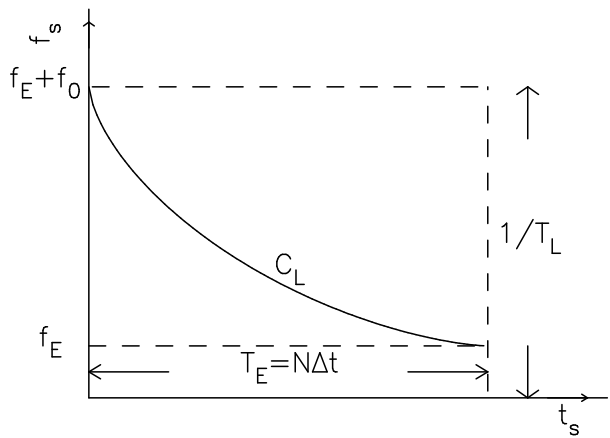}
\end{center}
\caption{The rate profile function,
Equation \ref{eq12}.}
\label{fig5}
\end{figure}

\begin{figure}[!htp]
\begin{center}
\includegraphics[height=15cm,angle=-90]{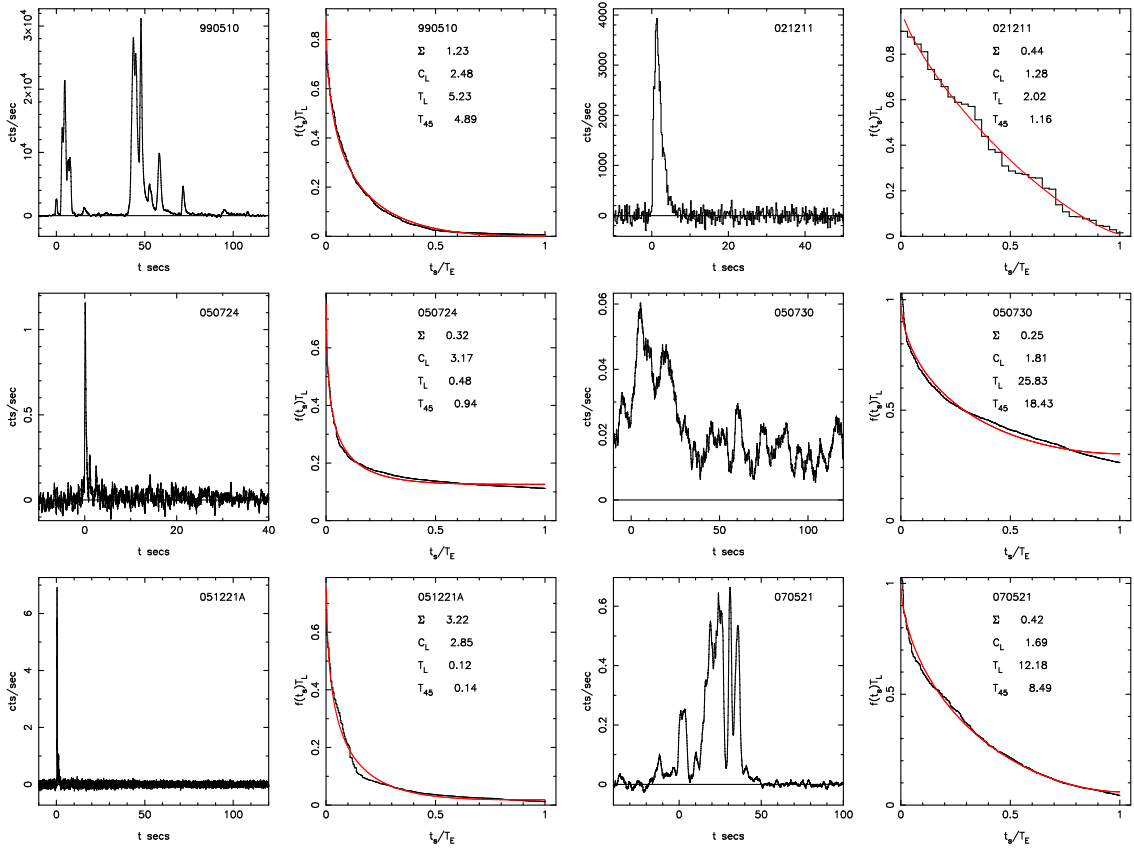}
\end{center}
\caption{Typical prompt emission light curves and the corresponding
rate profiles.
The light curves have been smoothed with a boxcar function of width
$T_{45}$ for display purposes.}
\label{fig6}
\end{figure}

\begin{figure}[!htp]
\begin{center}
\includegraphics[height=15cm,angle=-90]{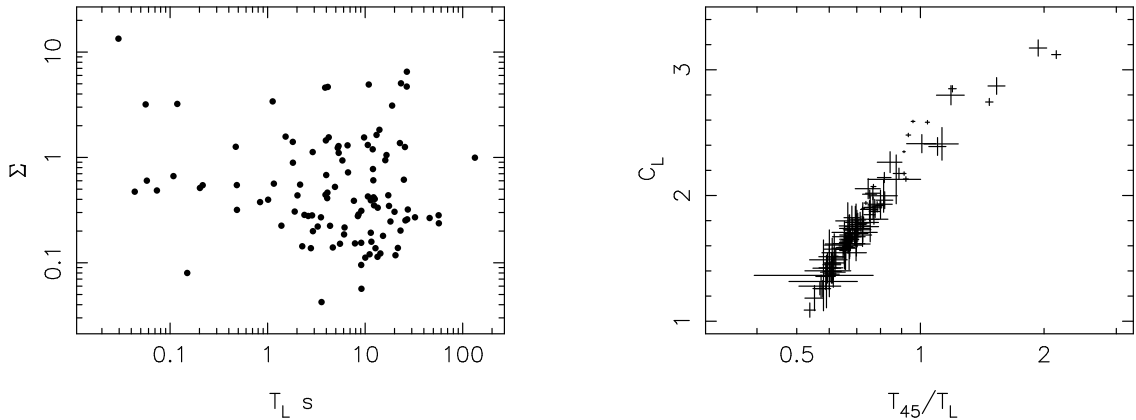}
\end{center}
\caption{Left-hand panel: The distribution of luminosity times, $T_{L}$ s,
and the $\Sigma$ statistic from the fit.
Right-hand panel: The correlation between rate profile index
$C_{L}$ and the ratio of $T_{45}$ derived directly from the data
and $T_{L}$ derived from the profile function fit.}
\label{fig7}
\end{figure}

\begin{figure}[!htp]
\begin{center}
\includegraphics[height=15cm,angle=-90]{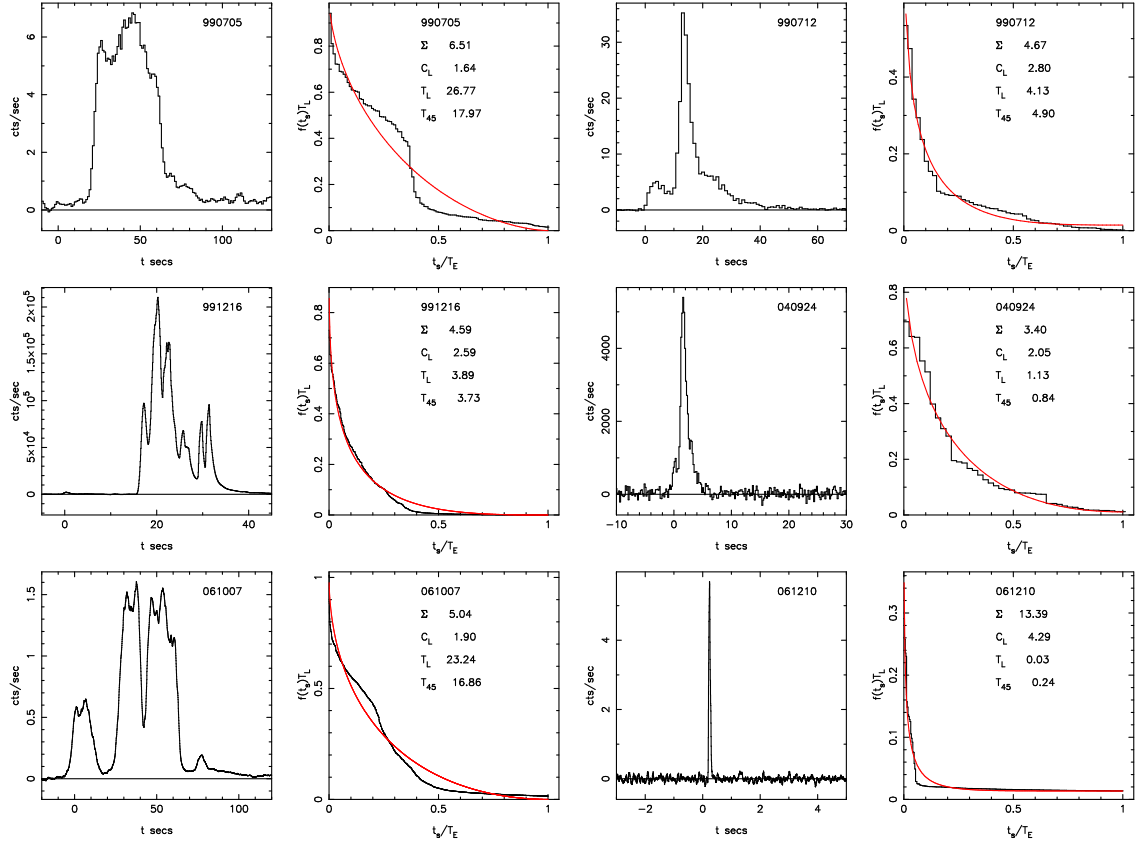}
\end{center}
\caption{Light curves and the corresponding rate profiles for which
the fit statistic $\Sigma$ is high.
The light curves have been smoothed with a boxcar function of width
$T_{45}$ for display purposes.}
\label{fig8}
\end{figure}

\begin{figure}[!htp]
\begin{center}
\includegraphics[height=15cm,angle=-90]{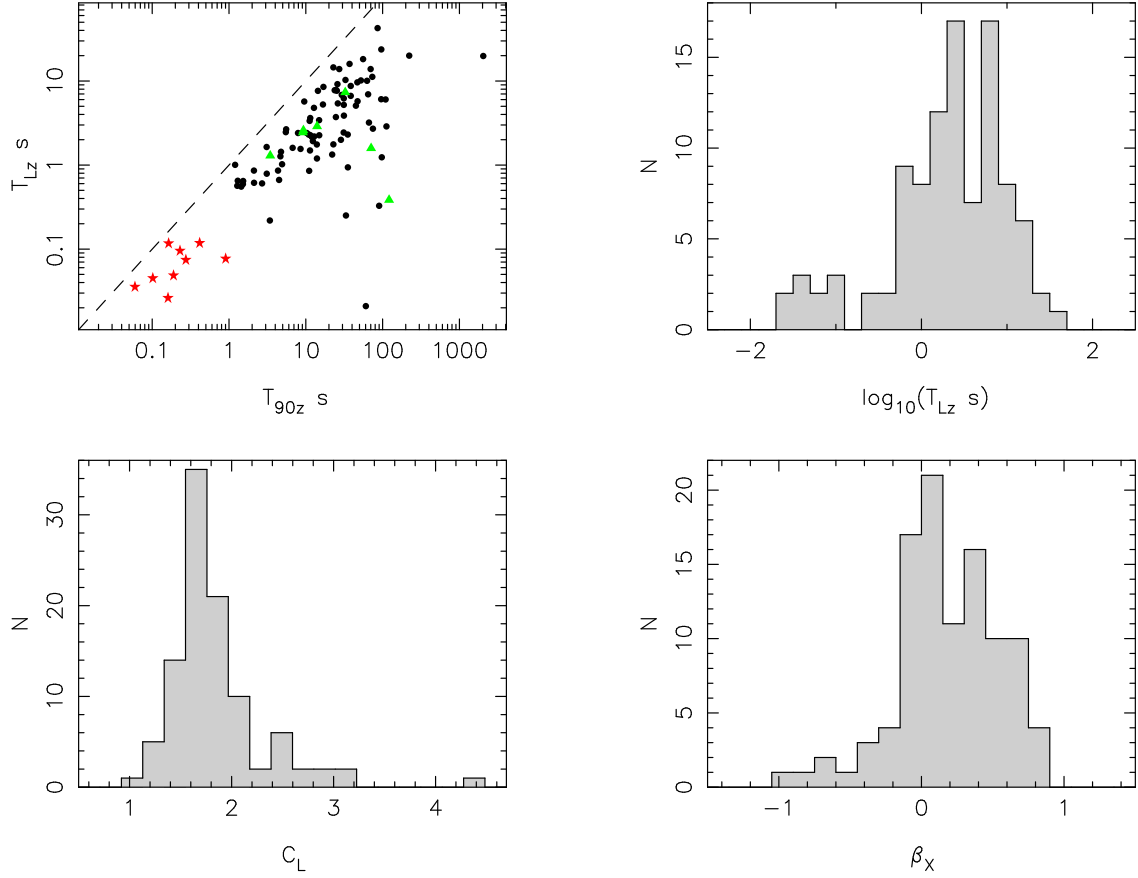}
\end{center}
\caption{Top panels: The correlation between luminosity time $T_{Lz}$ and
$T_{90z}$ and the distribution of $T_{Lz}$. The dashed line indicates
the equality $T_{Lz}=T_{90z}$.
Bottom panels: The distributions of $C_{L}$ and
$\beta_{X}$ for the GRBs in Table \ref{tab1}.}
\label{fig9}
\end{figure}

\begin{figure}[!htp]
\begin{center}
\includegraphics[height=15cm,angle=-90]{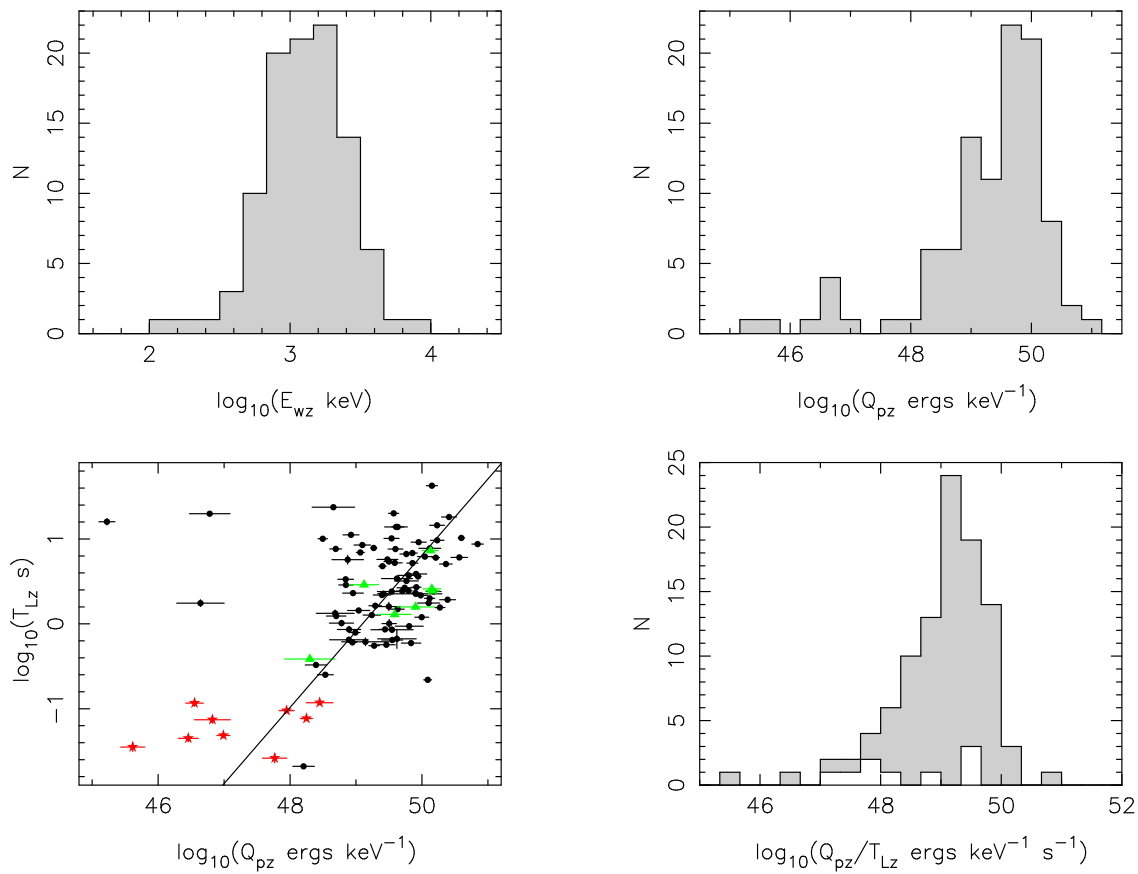}
\end{center}
\caption{Top panels: The distributions of $E_{wz}$ keV and $Q_{pz}$
ergs keV$^{-1}$.
Bottom panels: The correlation of $T_{Lz}$ s vs. $Q_{pz}$
and the
distribution of the peak luminosity density, $Q_{pz}/T_{Lz}$
ergs keV$^{-1}$ s$^{-1}$. The distribution for
short bursts is shown as the white histogram.}
\label{fig10}
\end{figure}

\begin{figure}[!htp]
\begin{center}
\includegraphics[height=15cm,angle=-90]{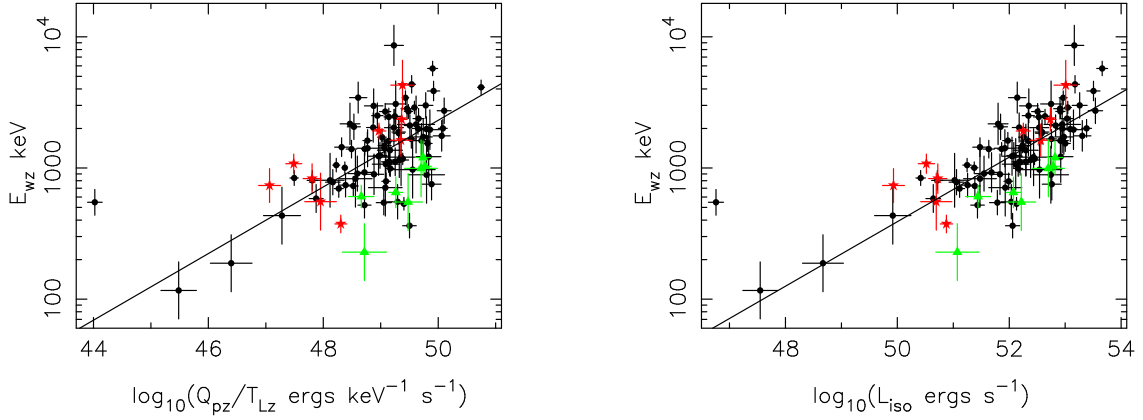}
\end{center}
\caption{Left-hand panel:
The correlation between characteristic energy $E_{wz}$ keV
and the peak luminosity density $Q_{pz}/T_{Lz}$ ergs keV$^{-1}$ 
s$^{-1}$.
Right-hand panel: $E_{wz}$ keV vs. the peak isotropic luminosity,
$L_{iso}=E_{wz}Q_{pz}/T_{Lz}$ ergs s$^{-1}$.
In both plots
solid dots for long GRBs, solid stars for short GRBs, solid triangles for
XRFs. The object with the largest $E_{wz}$, 8600 keV, is GRB050904.}
\label{fig11}
\end{figure}

\begin{figure}[!htp]
\begin{center}
\includegraphics[height=15cm,angle=-90]{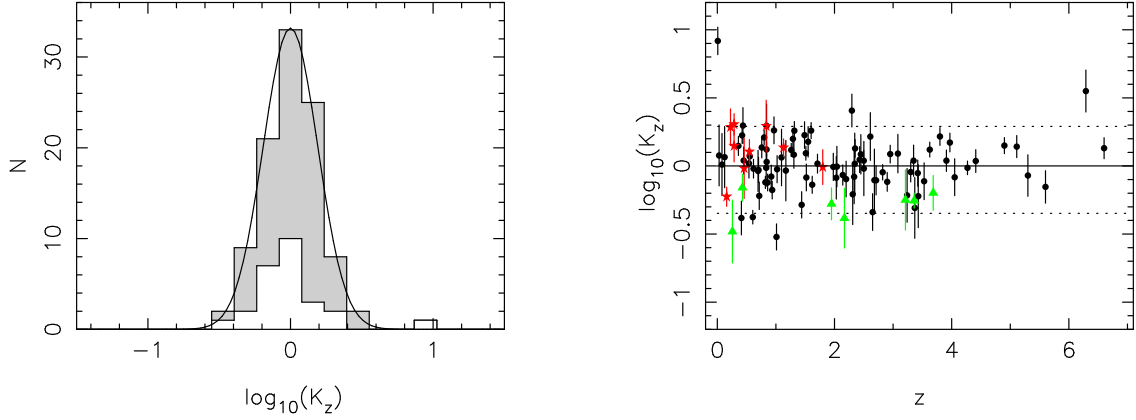}
\end{center}
\caption{The distribution of the photon energy/peak luminosity ratio $K_{z}$
corresponding to the scatter about the correlation line in the right-hand
panel of Figure \ref{fig11}. The white histogram shows the distribution
of the 26 pre-{\em Swift} bursts. The curve is the best fit Gaussian
distribution. The right-hand
panel shows $K_{z}$ vs. redshift $z$;
solid dots for long GRBs, solid stars for short GRBs, solid triangles for
XRFs. The horizontal dashed lines indicate the 90\% range. The objects
with unusually high $K_{z}$ are GRB980425/SN1998bw at low $z$ and
GRB050904 at high $z$.}
\label{fig12}
\end{figure}

\begin{figure}[!htp]
\begin{center}
\includegraphics[height=15cm,angle=-90]{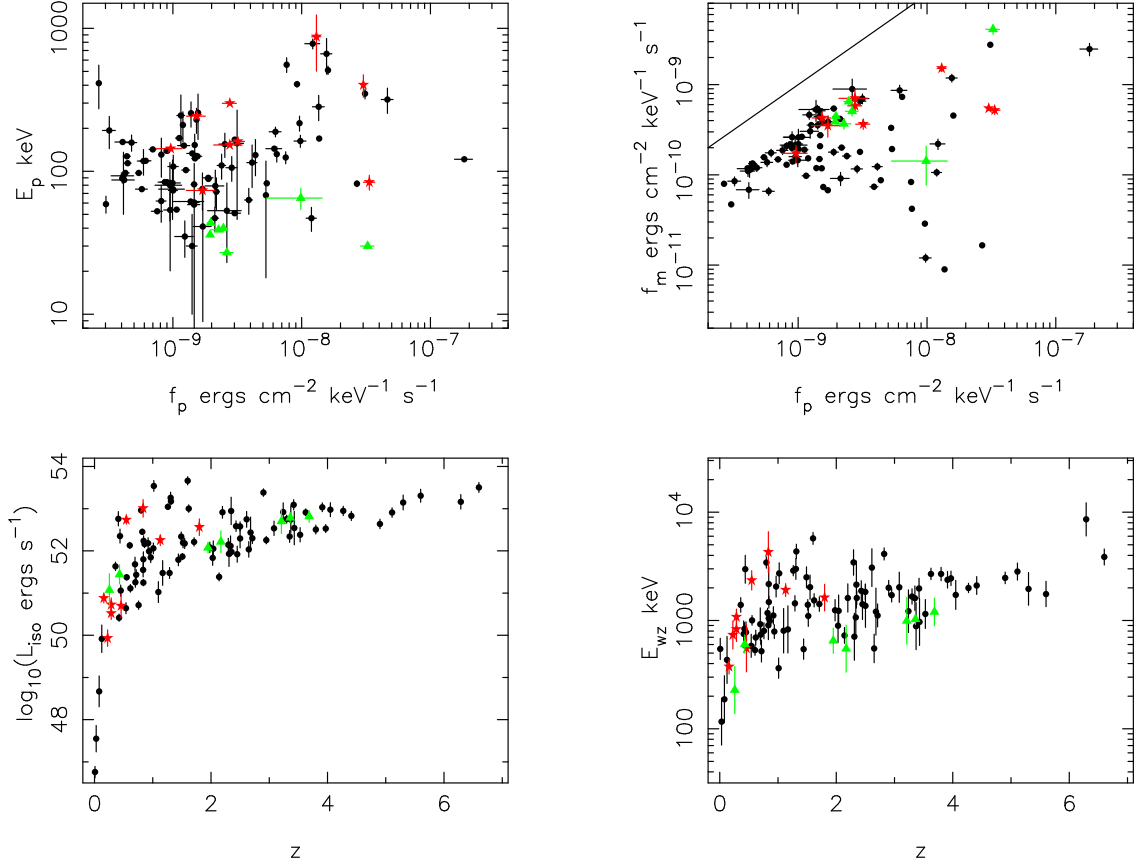}
\end{center}
\caption{Top panels: Observer frame parameters
$E_{p}$ and minimum observed flux density, $f_{m}$,
plotted against the peak flux density, $f_{p}$. The solid
line represents $f_{p}=f_{m}$.
Bottom panels: Source frame parameters, peak luminosity $L_{iso}$ and
characteristic energy $E_{wz}$ plotted vs. redshift $z$;
solid dots for long GRBs, solid stars for short GRBs, solid triangles for
XRFs.}
\label{fig13}
\end{figure}

$E_{wz}$ is a characteristic photon 
energy or a measure of the colour or hardness of the burst
and $Q_{pz}/T_{Lz}$ is a measure of the instantaneous peak brightness.
We have gathered and analysed
sufficient spectral and temporal data from 101 bursts to produce
the relation between $E_{wz}$ vs. $Q_{pz}/T_{Lz}$ and $E_{wz}$ vs. $L_{iso}$,
shown in Figure \ref{fig11},
which constitutes the closest thing we have to 
an intrinsic colour-magnitude diagram for the peak emission from GRBs,
$E_{wz}\propto L_{iso}^{0.25}$.
All bursts are clustered such that we can construct a intrinsic
colour-magnitude quasi constant $K_{z}$, which is a function of the 
source frame characteristic photon energy/peak luminosity ratio
given by Equation \ref{eq15}.
The range of equivalent isotropic energy that drives the expanding fireball
is very large, 6 orders of magnitude (Figure \ref{fig3}),
but the instantaneous hardness/brightness
of the peak emission covers a very small intrinsic dynamic range, $\approx4$.

The existence and form of $K_{z}$ indicates that the physical mechanism
for the Gamma-ray production at the photosphere of the fireball is
common to all bursts and is probably thermal although many other
possibilities are not ruled out. If the prompt spectra
are dominated by thermal photons the scatter in $K_{z}$ may be
attributed to variations in the size and/or Lorentz factor of the fireball.
XRFs have low $\Gamma_{0}$ and/or large radii. Short bursts
have high $\Gamma_{0}$ and/or small radii.
The relation between $T_{Lz}$ vs. $Q_{pz}$ clearly separates
short from long, but both classes have the same instantaneous peak
hardness/brightness. 

\acknowledgments

We gratefully acknowledge funding for {\em Swift} at the
University of Leicester by STFC. We thank the authors of the
{\em BATSE} ($cossc.gsfc.nasa.gov/docs/cgro/batse$),
{\em HETE} ($space.mit.edu/HETE$) and
{\em BeppoSAX} ($www.asdc.asi.it/grb\_wfc$) websites which gave us
access to the prompt lightcurves of pre-{\em Swift} bursts. We also thank
B. Zhang for valuable comments/discussions.





\clearpage



\newpage
\input{tab1}
\input{tab2}
\input{tab3}
\end{document}

%% file: tab1.tex
 \begin{deluxetable}{lrrrrrrrr}
 \tablewidth{0pt}
 \tablecaption{
 Observed spectral parameters;
 $\beta_{X}$, low spectral index,
 $E_{p}$, peak energy keV
 (if no error quoted then calculated assuming fixed
 $E_{cut}$, $E_{p}=(1-\beta_{X})*150$ keV),
 $\beta_{\gamma}$, high spectral index
 (if no error quoted then fixed at average of 1.3),
 $F_{tot}$, fluence ergs cm$^{-2}$,
 $E_{1}$ to $E_{2}$, observed energy band keV.
 References:
 (0) Swift BAT ibid.,
 (1) Golenetskii et al., 2005,
 (2) Villasenor et al., 2005,
 (3) Barthelmy et al. 2005, Amati et al. 2006,
     Campana et al. 2006,
 (4) Golenetskii et al., 2005,
 (5) Golenetskii et al. 2005, Burrows D.N. et al. 2006,
 (6) Romano et al., 2006,
 (7) Amati et al. 2007, Montanari \& Pian 2007,
 (8) Amati et al. 2007, Golenetskii et al. 2006,
 (9) Page et al. 2007,
 (10) Ghirlanda et al. 2004,
 (11) Friedman \& Bloom 2005,
 (12) Sakamoto et al. 2005,
 (13) Atteia et al. 2005,
 (14) Piro et al. 2005, Amati 2007,
 (15) Galassi et al. 2004, Amati 2007,
 (16) Schaefer 2007, Firmani et al. 2006.
 (17) Golenetskii et al. 2007.
 \label{tab1}}
 \tablehead{
 \colhead{GRB}&\colhead{$\beta_{X}$}&
 \colhead{$E_{p}$}&\colhead{$\beta_{\gamma}$}&
 \colhead{$F_{tot}$}&\colhead{$E_{1}$}&
 \colhead{$E_{2}$}&\colhead{refs}}
 \startdata
 970228&
 $ 0.54\pm 0.08$&
 $  115\pm   38$&
 $ 1.50\pm 0.40$&
 $(  11.0\pm 1.0)10^{-6}$&
      40&
     700&
 10
 \\
 970508&
 $ 0.71\pm 0.10$&
 $   79\pm   23$&
 $ 1.20\pm 0.25$&
 $(   1.8\pm 0.3)10^{-6}$&
      40&
     700&
 10
 \\
 971214&
 $-0.24\pm 0.10$&
 $  155\pm   30$&
 $ 1.70\pm 1.10$&
 $(   8.8\pm 0.9)10^{-6}$&
      40&
     700&
 10
 \\
 980425&
 $ 0.27\pm 0.13$&
 $  118\pm   24$&
 $ 1.30$&
 $(   3.8\pm 0.4)10^{-6}$&
      20&
    2000&
 10
 \\
 980613&
 $ 0.43\pm 0.20$&
 $   93\pm   43$&
 $ 1.70\pm 0.60$&
 $(   1.0\pm 0.2)10^{-6}$&
      40&
     700&
 10
 \\
 980703&
 $ 0.31\pm 0.14$&
 $  255\pm   51$&
 $ 1.39\pm 0.14$&
 $(  23.0\pm 0.2)10^{-6}$&
      20&
    2000&
 10
 \\
 990123&
 $-0.11\pm 0.08$&
 $  781\pm   62$&
 $ 1.45\pm 0.97$&
 $(   3.0\pm 0.4)10^{-4}$&
      40&
     700&
 10
 \\
 990506&
 $ 0.37\pm 0.15$&
 $  283\pm   57$&
 $ 1.15\pm 0.38$&
 $(   1.9\pm 0.2)10^{-4}$&
      20&
    2000&
 10
 \\
 990510&
 $ 0.23\pm 0.05$&
 $  163\pm   16$&
 $ 1.70\pm 0.40$&
 $(   1.9\pm 0.2)10^{-5}$&
      40&
     700&
 10
 \\
 990705&
 $ 0.05\pm 0.20$&
 $  189\pm   15$&
 $ 1.20\pm 0.10$&
 $(   7.5\pm 0.8)10^{-5}$&
      40&
     700&
 10
 \\
 990712&
 $ 0.88\pm 0.07$&
 $   65\pm   11$&
 $ 1.48\pm 0.56$&
 $(   0.6\pm 0.3)10^{-5}$&
      40&
     700&
 10
 \\
 991216&
 $ 0.23\pm 0.13$&
 $  318\pm   64$&
 $ 1.18\pm 0.39$&
 $(   1.9\pm 0.2)10^{-4}$&
      20&
    2000&
 10
 \\
 010921&
 $ 0.49\pm 0.16$&
 $  106\pm   21$&
 $ 1.30$&
 $(  10.0\pm 1.0)10^{-6}$&
      30&
     700&
 10
 \\
 011121&
 $ 0.42\pm 0.14$&
 $  217\pm   26$&
 $ 1.30$&
 $(  96.6\pm 1.0)10^{-6}$&
      40&
     700&
 14
 \\
 011211&
 $-0.16\pm 0.09$&
 $   59\pm    8$&
 $ 1.30$&
 $(   5.1\pm 0.2)10^{-6}$&
      40&
     700&
 14
 \\
 021004&
 $ 0.01\pm 0.19$&
 $   80\pm   35$&
 $ 1.30\pm 0.46$&
 $(   2.6\pm 0.6)10^{-6}$&
       2&
     400&
 11
 \\
 021211&
 $-0.15\pm 0.09$&
 $   47\pm    9$&
 $ 1.37\pm 0.42$&
 $(   2.2\pm 0.2)10^{-6}$&
      30&
     400&
 10
 \\
 030115A&
 $ 0.28\pm 0.14$&
 $   83\pm   37$&
 $ 1.20\pm 0.40$&
 $(   2.3\pm 0.3)10^{-6}$&
       2&
     400&
 16
 \\
 030226&
 $-0.05\pm 0.10$&
 $  108\pm   22$&
 $ 1.30$&
 $(   6.4\pm 0.6)10^{-6}$&
      30&
     400&
 10
 \\
 030323&
 $-0.20\pm 0.20$&
 $   53\pm   30$&
 $ 1.30$&
 $(   1.2\pm 0.3)10^{-6}$&
       2&
     400&
 13
 \\
 030328&
 $ 0.00\pm 0.11$&
 $  110\pm   22$&
 $ 1.30$&
 $(   2.6\pm 0.2)10^{-5}$&
      30&
     400&
 10
 \\
 030429&
 $ 0.10\pm 0.20$&
 $   35\pm   10$&
 $ 1.30$&
 $(   0.8\pm 0.1)10^{-6}$&
       2&
     400&
 12
 \\
 040924&
 $ 0.17\pm 0.05$&
 $  125\pm   12$&
 $ 1.30$&
 $(   2.7\pm 0.1)10^{-6}$&
      20&
     500&
 16
 \\
 041006&
 $ 0.37\pm 0.10$&
 $   63\pm   13$&
 $ 1.30$&
 $(   7.0\pm 0.5)10^{-6}$&
      30&
     400&
 15
 \\
 050126&
 $ 0.06\pm 0.20$&
 $  158\pm   20$&
 $ 1.30$&
 $(   8.4\pm 0.8)10^{-7}$&
      15&
     150&
  0
 \\
 050315&
 $ 0.76\pm 0.06$&
 $   36$&
 $ 1.30$&
 $(   3.2\pm 0.1)10^{-6}$&
      15&
     150&
  0
 \\
 050318&
 $ 0.34\pm 0.20$&
 $   47\pm    9$&
 $ 1.30$&
 $(  10.8\pm 0.8)10^{-7}$&
      15&
     150&
  0
 \\
 050319&
 $ 0.66\pm 0.15$&
 $   51$&
 $ 1.30$&
 $(   1.3\pm 0.1)10^{-6}$&
      15&
     150&
  0
 \\
 050401&
 $ 0.11\pm 0.07$&
 $  132\pm   16$&
 $ 1.30$&
 $(   8.2\pm 0.3)10^{-6}$&
      15&
     150&
  0
 \\
 050505&
 $-0.01\pm 0.20$&
 $  102\pm    0$&
 $ 1.30$&
 $(   2.5\pm 0.2)10^{-6}$&
      15&
     150&
  0
 \\
 050509B&
 $ 0.04\pm 0.20$&
 $  144$&
 $ 1.30$&
 $(   0.9\pm 0.2)10^{-8}$&
      15&
     150&
  0
 \\
 050525A&
 $-0.13\pm 0.07$&
 $   82\pm    4$&
 $ 1.30$&
 $(  15.3\pm 0.2)10^{-6}$&
      15&
     150&
  0
 \\
 050603&
 $-0.19\pm 0.04$&
 $  349\pm   28$&
 $ 1.30$&
 $(   6.4\pm 0.2)10^{-6}$&
      15&
     150&
  1
 \\
 050709&
 $-0.47\pm 0.13$&
 $   83\pm   10$&
 $ 1.30$&
 $(   4.0\pm 0.4)10^{-7}$&
       2&
     400&
  2
 \\
 050724&
 $ 0.80\pm 0.17$&
 $   30$&
 $ 1.30$&
 $(   1.0\pm 0.1)10^{-6}$&
      15&
     150&
  3
 \\
 050730&
 $ 0.15\pm 0.10$&
 $  127$&
 $ 1.30$&
 $(   2.4\pm 0.2)10^{-6}$&
      15&
     150&
  0
 \\
 050802&
 $ 0.15\pm 0.10$&
 $  127$&
 $ 1.30$&
 $(   2.0\pm 0.2)10^{-6}$&
      15&
     150&
  0
 \\
 050803&
 $ 0.05\pm 0.08$&
 $  142$&
 $ 1.30$&
 $(   2.2\pm 0.1)10^{-6}$&
      15&
     150&
  0
 \\
 050813&
 $-0.02\pm 0.20$&
 $  153$&
 $ 1.30$&
 $(   0.4\pm 0.1)10^{-7}$&
      15&
     150&
  0
 \\
 050814&
 $ 0.61\pm 0.13$&
 $   58$&
 $ 1.30$&
 $(   2.0\pm 0.2)10^{-6}$&
      15&
     150&
  0
 \\
 050820A&
 $-0.03\pm 0.09$&
 $  246\pm   96$&
 $ 1.30$&
 $(   3.4\pm 0.2)10^{-6}$&
      15&
     150&
  0
 \\
 050904&
 $ 0.07\pm 0.14$&
 $  413\pm  140$&
 $ 1.30$&
 $(   4.8\pm 0.2)10^{-6}$&
      15&
     150&
  0
 \\
 050908&
 $ 0.50\pm 0.12$&
 $   75$&
 $ 1.30$&
 $(   4.8\pm 0.5)10^{-7}$&
      15&
     150&
  0
 \\
 050922C&
 $ 0.04\pm 0.05$&
 $  130\pm   37$&
 $ 1.30$&
 $(  16.2\pm 0.5)10^{-7}$&
      15&
     150&
  0
 \\
 051022&
 $ 0.18\pm 0.04$&
 $  510\pm   35$&
 $ 1.30$&
 $(  26.1\pm 0.9)10^{-5}$&
      20&
    2000&
  4
 \\
 051109A&
 $ 0.23\pm 0.15$&
 $  157\pm  111$&
 $ 1.30$&
 $(   2.2\pm 0.3)10^{-6}$&
      15&
     150&
  4
 \\
 051109B&
 $-0.11\pm 0.20$&
 $   41\pm   32$&
 $ 1.30$&
 $(   2.6\pm 0.4)10^{-7}$&
      15&
     150&
  0
 \\
 051111&
 $ 0.04\pm 0.05$&
 $  211\pm   50$&
 $ 1.30$&
 $(   4.1\pm 0.1)10^{-6}$&
      15&
     150&
  0
 \\
 051221A&
 $ 0.08\pm 0.04$&
 $  402\pm   72$&
 $ 1.30$&
 $(  11.5\pm 0.3)10^{-7}$&
      15&
     150&
  5
 \\
 060108&
 $ 0.65\pm 0.12$&
 $   52$&
 $ 1.30$&
 $(   3.7\pm 0.4)10^{-7}$&
      15&
     150&
  0
 \\
 060115&
 $ 0.09\pm 0.20$&
 $   62\pm   18$&
 $ 1.30$&
 $(   1.7\pm 0.2)10^{-6}$&
      15&
     150&
  0
 \\
 060116&
 $-0.01\pm 0.14$&
 $  151$&
 $ 1.30$&
 $(   2.4\pm 0.3)10^{-6}$&
      15&
     150&
  0
 \\
 060124&
 $ 0.66\pm 0.17$&
 $  193\pm   49$&
 $ 1.30$&
 $(   4.6\pm 0.5)10^{-7}$&
      15&
     150&
  6
 \\
 060206&
 $ 0.38\pm 0.19$&
 $   75\pm   22$&
 $ 1.30$&
 $(   8.3\pm 0.4)10^{-7}$&
      15&
     150&
  0
 \\
 060210&
 $ 0.18\pm 0.12$&
 $  123$&
 $ 1.30$&
 $(   7.7\pm 0.4)10^{-6}$&
      15&
     150&
  0
 \\
 060218&
 $-0.76\pm 0.20$&
 $   30\pm   20$&
 $ 1.30$&
 $(   1.6\pm 0.2)10^{-6}$&
      15&
     150&
  7
 \\
 060223A&
 $ 0.44\pm 0.09$&
 $   83$&
 $ 1.30$&
 $(   6.7\pm 0.5)10^{-7}$&
      15&
     150&
  0
 \\
 060418&
 $ 0.38\pm 0.04$&
 $  230\pm   46$&
 $ 1.30$&
 $(   8.3\pm 0.3)10^{-6}$&
      15&
     150&
  0
 \\
 060502A&
 $ 0.07\pm 0.05$&
 $  139$&
 $ 1.30$&
 $(   2.3\pm 0.1)10^{-6}$&
      15&
     150&
  0
 \\
 060502B&
 $-0.08\pm 0.20$&
 $  162$&
 $ 1.30$&
 $(   4.0\pm 0.5)10^{-8}$&
      15&
     150&
  0
 \\
 060510B&
 $ 0.48\pm 0.20$&
 $   89\pm    6$&
 $ 1.30$&
 $(   4.1\pm 0.2)10^{-6}$&
      15&
     150&
  0
 \\
 060522&
 $ 0.21\pm 0.12$&
 $  118$&
 $ 1.30$&
 $(   1.1\pm 0.1)10^{-6}$&
      15&
     150&
  0
 \\
 060526&
 $ 0.74\pm 0.16$&
 $   39$&
 $ 1.30$&
 $(   1.3\pm 0.2)10^{-6}$&
      15&
     150&
  0
 \\
 060604&
 $ 0.59\pm 0.20$&
 $   61$&
 $ 1.30$&
 $(   0.4\pm 0.1)10^{-6}$&
      15&
     150&
  0
 \\
 060605&
 $-0.07\pm 0.11$&
 $  160$&
 $ 1.30$&
 $(   7.0\pm 0.9)10^{-7}$&
      15&
     150&
  0
 \\
 060607A&
 $ 0.05\pm 0.20$&
 $  131\pm   40$&
 $ 1.30$&
 $(   2.6\pm 0.1)10^{-6}$&
      15&
     150&
  0
 \\
 060614&
 $ 0.66\pm 0.03$&
 $   68\pm   50$&
 $ 1.30$&
 $(  20.4\pm 0.4)10^{-6}$&
      15&
     150&
  8
 \\
 060707&
 $-0.42\pm 0.20$&
 $   60\pm   53$&
 $ 1.30$&
 $(   1.6\pm 0.2)10^{-6}$&
      15&
     150&
  0
 \\
 060714&
 $ 0.64\pm 0.08$&
 $   54$&
 $ 1.30$&
 $(   2.8\pm 0.2)10^{-6}$&
      15&
     150&
  0
 \\
 060729&
 $ 0.51\pm 0.11$&
 $   73$&
 $ 1.30$&
 $(   2.6\pm 0.2)10^{-6}$&
      15&
     150&
  0
 \\
 060801&
 $-0.99\pm 0.19$&
 $  298$&
 $ 1.30$&
 $(   8.0\pm 1.0)10^{-8}$&
      15&
     150&
  0
 \\
 060814&
 $ 0.43\pm 0.16$&
 $  257\pm   90$&
 $ 1.30$&
 $(  14.6\pm 0.2)10^{-6}$&
      15&
     150&
  0
 \\
 060904B&
 $ 0.40\pm 0.11$&
 $   90$&
 $ 1.30$&
 $(   1.6\pm 0.1)10^{-6}$&
      15&
     150&
  0
 \\
 060906&
 $ 0.71\pm 0.08$&
 $   43$&
 $ 1.30$&
 $(   2.2\pm 0.1)10^{-6}$&
      15&
     150&
  0
 \\
 060908&
 $-0.15\pm 0.16$&
 $  153\pm   29$&
 $ 1.30$&
 $(   2.8\pm 0.1)10^{-6}$&
      15&
     150&
  0
 \\
 060912&
 $ 0.45\pm 0.06$&
 $   82$&
 $ 1.30$&
 $(  13.5\pm 0.6)10^{-7}$&
      15&
     150&
  0
 \\
 060927&
 $-0.07\pm 0.20$&
 $   72\pm   18$&
 $ 1.30$&
 $(  11.3\pm 0.7)10^{-7}$&
      15&
     150&
  0
 \\
 061004&
 $ 0.46\pm 0.07$&
 $   80$&
 $ 1.30$&
 $(   5.7\pm 0.3)10^{-7}$&
      15&
     150&
  0
 \\
 061006&
 $-0.38\pm 0.20$&
 $  664\pm  186$&
 $ 1.30$&
 $(   1.4\pm 0.1)10^{-6}$&
      15&
     150&
 17
 \\
 061007&
 $-0.38\pm 0.02$&
 $  407\pm   19$&
 $ 1.30$&
 $(  44.4\pm 0.6)10^{-6}$&
      15&
     150&
  0
 \\
 061110A&
 $ 0.35\pm 0.09$&
 $   97$&
 $ 1.30$&
 $(  10.6\pm 0.8)10^{-7}$&
      15&
     150&
  0
 \\
 061121&
 $ 0.05\pm 0.02$&
 $  557\pm   66$&
 $ 1.30$&
 $(  13.7\pm 0.2)10^{-6}$&
      15&
     150&
  9
 \\
 061201&
 $-0.67\pm 0.15$&
 $  873\pm  371$&
 $ 1.30$&
 $(   3.3\pm 0.3)10^{-7}$&
      15&
     150&
  0
 \\
 061210&
 $ 0.19\pm 0.20$&
 $  121$&
 $ 1.30$&
 $(   1.1\pm 0.2)10^{-6}$&
      15&
     150&
  0
 \\
 061217&
 $-0.62\pm 0.20$&
 $  243$&
 $ 1.30$&
 $(   4.2\pm 0.7)10^{-8}$&
      15&
     150&
  0
 \\
 061222B&
 $ 0.71\pm 0.20$&
 $   40\pm    2$&
 $ 1.30$&
 $(   2.2\pm 0.2)10^{-6}$&
      15&
     150&
  0
 \\
 070110&
 $ 0.24\pm 0.08$&
 $  114$&
 $ 1.30$&
 $(   1.6\pm 0.1)10^{-6}$&
      15&
     150&
  0
 \\
 070208&
 $ 0.50\pm 0.20$&
 $   75$&
 $ 1.30$&
 $(   0.4\pm 0.1)10^{-6}$&
      15&
     150&
  0
 \\
 070318&
 $ 0.07\pm 0.06$&
 $  139$&
 $ 1.30$&
 $(   2.5\pm 0.1)10^{-6}$&
      15&
     150&
  0
 \\
 070411&
 $ 0.35\pm 0.07$&
 $   97$&
 $ 1.30$&
 $(   2.7\pm 0.2)10^{-6}$&
      15&
     150&
  0
 \\
 070506&
 $-0.11\pm 0.20$&
 $   53\pm   33$&
 $ 1.30$&
 $(   2.1\pm 0.2)10^{-7}$&
      15&
     150&
  0
 \\
 070508&
 $-0.13\pm 0.02$&
 $  169$&
 $ 1.30$&
 $(  19.6\pm 0.3)10^{-6}$&
      15&
     150&
  0
 \\
 070521&
 $-0.11\pm 0.03$&
 $  166$&
 $ 1.30$&
 $(   8.0\pm 0.2)10^{-6}$&
      15&
     150&
  0
 \\
 070529&
 $ 0.11\pm 0.12$&
 $  133$&
 $ 1.30$&
 $(   2.6\pm 0.2)10^{-6}$&
      15&
     150&
  0
 \\
 070611&
 $ 0.44\pm 0.18$&
 $   83$&
 $ 1.30$&
 $(   3.9\pm 0.6)10^{-7}$&
      15&
     150&
  0
 \\
 070612A&
 $ 0.41\pm 0.07$&
 $   88$&
 $ 1.30$&
 $(  10.6\pm 0.6)10^{-6}$&
      15&
     150&
  0
 \\
 070714B&
 $ 0.04\pm 0.15$&
 $  144$&
 $ 1.30$&
 $(   6.4\pm 0.9)10^{-7}$&
      15&
     150&
  0
 \\
 070721B&
 $-0.14\pm 0.08$&
 $  171$&
 $ 1.30$&
 $(   3.0\pm 0.2)10^{-6}$&
      15&
     150&
  0
 \\
 070724A&
 $ 0.51\pm 0.20$&
 $   73$&
 $ 1.30$&
 $(   2.8\pm 0.7)10^{-8}$&
      15&
     150&
  0
 \\
 070802&
 $ 0.42\pm 0.18$&
 $   87$&
 $ 1.30$&
 $(   2.5\pm 0.5)10^{-7}$&
      15&
     150&
  0
 \\
 070810A&
 $ 0.82\pm 0.10$&
 $   26$&
 $ 1.30$&
 $(   6.1\pm 0.7)10^{-7}$&
      15&
     150&
  0
 \\
 \enddata
 \end{deluxetable}

%% file: tab2.tex
 \begin{deluxetable}{lllrrrr}
 \tablewidth{0pt}
 \tablecaption{
 Observed temporal parameters;
 $T_{90}$ secs,
 luminosity time $T_{L}$ secs,
 luminosity index $C_{L}$ ,
 luminosity profile fitting statistic $\Sigma$.
 Errors were estimated by assuming $\Sigma$ was distributed
 as $\chi^{2}$.
 \label{tab2}}
 \tablehead{
 \colhead{GRB}&\colhead{instr}&
 \colhead{class}&\colhead{$T_{90}$}&
 \colhead{$T_{L}$}&\colhead{$C_{L}$}&\colhead{$\Sigma$}}
 \startdata
 970228&
 SAX&
 Long&
  80.0&
 $  9.7\pm  0.6$&
 $2.41\pm0.07$&
 1.55
 \\
 970508&
 SAX&
 Long&
  20.0&
 $  4.2\pm  0.1$&
 $2.01\pm0.03$&
 1.55
 \\
 971214&
 SAX&
 Long&
  35.0&
 $ 10.6\pm  0.2$&
 $1.52\pm0.02$&
 1.31
 \\
 980425&
 SAX&
 Long&
  37.4&
 $  16.\pm   1.$&
 $1.54\pm0.13$&
 0.94
 \\
 980613&
 BATSE&
 Long&
  20.0&
 $  12.\pm   1.$&
 $1.32\pm0.21$&
 1.19
 \\
 980703&
 BATSE&
 Long&
   102&
 $ 20.0\pm  0.4$&
 $1.56\pm0.03$&
 0.30
 \\
 990123&
 SAX&
 Long&
 100.0&
 $ 22.7\pm  0.2$&
 $1.94\pm0.01$&
 1.37
 \\
 990506&
 BATSE&
 Long&
   220&
 $14.00\pm 0.10$&
 $2.35\pm0.01$&
 1.83
 \\
 990510&
 SAX&
 Long&
  75.0&
 $ 5.23\pm 0.06$&
 $2.48\pm0.01$&
 1.23
 \\
 990705&
 SAX&
 Long&
  42.0&
 $ 26.8\pm  0.9$&
 $1.64\pm0.03$&
 6.51
 \\
 990712&
 SAX&
 XRF&
  20.0&
 $  4.1\pm  0.3$&
 $2.80\pm0.08$&
 4.67
 \\
 991216&
 BATSE&
 Long&
  24.9&
 $ 3.89\pm 0.04$&
 $2.59\pm0.01$&
 4.59
 \\
 010921&
 HETE-2&
 Long&
  24.6&
 $ 12.3\pm  0.4$&
 $1.26\pm0.05$&
 0.35
 \\
 011121&
 SAX&
 Long&
  37.0&
 $ 18.9\pm  0.6$&
 $2.14\pm0.04$&
 3.10
 \\
 011211&
 SAX&
 Long&
   270&
 $ 133.\pm   5.$&
 $1.26\pm0.06$&
 0.99
 \\
 021004&
 HETE-2&
 Long&
  49.7&
 $ 11.4\pm  0.7$&
 $1.62\pm0.10$&
 0.39
 \\
 021211&
 HETE-2&
 Long&
   2.4&
 $  2.0\pm  0.2$&
 $1.28\pm0.09$&
 0.44
 \\
 030115A&
 HETE-2&
 Long&
  36.0&
 $  8.4\pm  0.4$&
 $1.73\pm0.08$&
 0.28
 \\
 030226&
 HETE-2&
 Long&
  76.8&
 $ 27.3\pm  0.9$&
 $1.60\pm0.06$&
 0.32
 \\
 030323&
 HETE-2&
 Long&
  19.6&
 $  2.9\pm  0.8$&
 $1.36\pm0.28$&
 0.20
 \\
 030328&
 HETE-2&
 Long&
   140&
 $ 45.8\pm  0.7$&
 $1.66\pm0.02$&
 0.27
 \\
 030429&
 HETE-2&
 Long&
  24.6&
 $  5.8\pm  0.7$&
 $2.13\pm0.19$&
 0.94
 \\
 040924&
 HETE-2&
 Long&
   5.0&
 $ 1.13\pm 0.07$&
 $2.05\pm0.08$&
 3.40
 \\
 041006&
 HETE-2&
 Long&
  24.6&
 $ 13.1\pm  0.4$&
 $1.37\pm0.04$&
 1.63
 \\
 050126&
 Swift&
 Long&
  25.7&
 $  7.7\pm  0.4$&
 $1.65\pm0.09$&
 0.39
 \\
 050315&
 Swift&
 XRF&
  96.0&
 $ 21.6\pm  0.5$&
 $1.62\pm0.04$&
 0.14
 \\
 050318&
 Swift&
 Long&
  31.3&
 $  5.3\pm  0.3$&
 $1.40\pm0.09$&
 1.11
 \\
 050319&
 Swift&
 Long&
   149&
 $  4.0\pm  0.3$&
 $1.93\pm0.11$&
 0.68
 \\
 050401&
 Swift&
 Long&
  33.3&
 $  6.1\pm  0.2$&
 $1.78\pm0.05$&
 0.19
 \\
 050505&
 Swift&
 Long&
  63.0&
 $ 11.4\pm  0.4$&
 $1.50\pm0.08$&
 0.19
 \\
 050509B&
 Swift&
 Short&
  0.07&
 $0.043\pm0.005$&
 $1.61\pm0.20$&
 0.47
 \\
 050525A&
 Swift&
 Long&
   8.8&
 $  3.9\pm  0.1$&
 $1.58\pm0.03$&
 1.45
 \\
 050603&
 Swift&
 Long&
  13.0&
 $ 0.84\pm 0.02$&
 $2.74\pm0.03$&
 0.38
 \\
 050709&
 HETE-2&
 Short&
  0.22&
 $0.056\pm0.004$&
 $2.26\pm0.08$&
 3.19
 \\
 050724&
 Swift&
 XRF&
   152&
 $ 0.48\pm 0.02$&
 $3.17\pm0.06$&
 0.32
 \\
 050730&
 Swift&
 Long&
   155&
 $ 25.8\pm  0.8$&
 $1.81\pm0.06$&
 0.25
 \\
 050802&
 Swift&
 Long&
  30.9&
 $  6.1\pm  0.3$&
 $1.59\pm0.09$&
 0.22
 \\
 050803&
 Swift&
 Long&
  89.0&
 $ 14.3\pm  0.5$&
 $1.85\pm0.05$&
 0.12
 \\
 050813&
 Swift&
 Short&
  0.45&
 $0.073\pm0.009$&
 $1.71\pm0.23$&
 0.49
 \\
 050814&
 Swift&
 Long&
   144&
 $ 11.1\pm  0.6$&
 $1.74\pm0.11$&
 0.12
 \\
 050820A&
 Swift&
 Long&
   240&
 $ 11.6\pm  0.2$&
 $1.75\pm0.03$&
 0.16
 \\
 050904&
 Swift&
 Long&
   173&
 $  57.\pm   1.$&
 $1.62\pm0.04$&
 0.28
 \\
 050908&
 Swift&
 Long&
  20.3&
 $  5.5\pm  0.3$&
 $1.45\pm0.09$&
 0.15
 \\
 050922C&
 Swift&
 Long&
   4.1&
 $ 1.81\pm 0.07$&
 $1.65\pm0.05$&
 0.89
 \\
 051022&
 Konus&
 Long&
   197&
 $ 10.9\pm  0.3$&
 $1.81\pm0.04$&
 4.91
 \\
 051109A&
 Swift&
 Long&
  37.0&
 $  2.9\pm  0.2$&
 $2.00\pm0.13$&
 0.28
 \\
 051109B&
 Swift&
 Long&
  15.0&
 $  1.9\pm  0.2$&
 $1.79\pm0.16$&
 0.31
 \\
 051111&
 Swift&
 Long&
  42.6&
 $ 13.4\pm  0.3$&
 $1.66\pm0.04$&
 0.11
 \\
 051221A&
 Swift&
 Short&
  1.40&
 $0.118\pm0.003$&
 $2.85\pm0.03$&
 3.22
 \\
 060108&
 Swift&
 Long&
  14.4&
 $  4.4\pm  0.3$&
 $1.46\pm0.11$&
 0.22
 \\
 060115&
 Swift&
 Long&
   141&
 $ 17.5\pm  0.6$&
 $1.78\pm0.06$&
 0.35
 \\
 060116&
 Swift&
 Long&
   105&
 $  9.1\pm  0.5$&
 $1.57\pm0.12$&
 0.15
 \\
 060124&
 Swift&
 Long&
   321&
 $  4.1\pm  0.3$&
 $1.58\pm0.13$&
 0.41
 \\
 060206&
 Swift&
 Long&
   7.7&
 $  3.3\pm  0.1$&
 $1.42\pm0.05$&
 0.22
 \\
 060210&
 Swift&
 Long&
   220&
 $ 24.9\pm  0.6$&
 $1.92\pm0.04$&
 0.61
 \\
 060218&
 Swift&
 Long&
  2100&
 $ 20.4\pm  0.9$&
 $1.82\pm0.09$&
 0.12
 \\
 060223A&
 Swift&
 Long&
  11.4&
 $  4.7\pm  0.3$&
 $1.36\pm0.10$&
 0.14
 \\
 060418&
 Swift&
 Long&
  95.8&
 $ 16.5\pm  0.3$&
 $1.93\pm0.02$&
 1.05
 \\
 060502A&
 Swift&
 Long&
  32.0&
 $ 12.0\pm  0.4$&
 $1.38\pm0.06$&
 0.77
 \\
 060502B&
 Swift&
 Short&
  0.13&
 $0.058\pm0.006$&
 $2.41\pm0.13$&
 0.60
 \\
 060510B&
 Swift&
 Long&
   276&
 $  57.\pm   1.$&
 $1.65\pm0.04$&
 0.24
 \\
 060522&
 Swift&
 Long&
  69.3&
 $  9.1\pm  0.4$&
 $1.77\pm0.09$&
 0.10
 \\
 060526&
 Swift&
 XRF&
   298&
 $  6.7\pm  0.3$&
 $2.39\pm0.07$&
 0.72
 \\
 060604&
 Swift&
 Long&
   7.8&
 $  2.3\pm  0.2$&
 $1.49\pm0.24$&
 0.14
 \\
 060605&
 Swift&
 Long&
  14.8&
 $  7.9\pm  0.4$&
 $1.46\pm0.09$&
 0.15
 \\
 060607A&
 Swift&
 Long&
   100&
 $ 15.2\pm  0.4$&
 $1.88\pm0.04$&
 0.18
 \\
 060614&
 Swift&
 Long&
   108&
 $ 26.7\pm  0.4$&
 $2.13\pm0.02$&
 4.70
 \\
 060707&
 Swift&
 Long&
  66.2&
 $ 10.0\pm  0.5$&
 $1.72\pm0.09$&
 0.11
 \\
 060714&
 Swift&
 Long&
   115&
 $ 23.0\pm  0.6$&
 $1.62\pm0.05$&
 0.20
 \\
 060729&
 Swift&
 Long&
   112&
 $ 17.2\pm  0.5$&
 $1.59\pm0.05$&
 0.44
 \\
 060801&
 Swift&
 Short&
  0.49&
 $ 0.20\pm 0.01$&
 $1.70\pm0.11$&
 0.51
 \\
 060814&
 Swift&
 Long&
   128&
 $ 25.5\pm  0.3$&
 $2.17\pm0.01$&
 1.25
 \\
 060904B&
 Swift&
 Long&
   190&
 $  4.9\pm  0.2$&
 $2.17\pm0.04$&
 0.53
 \\
 060906&
 Swift&
 XRF&
  43.5&
 $ 12.1\pm  0.4$&
 $1.67\pm0.07$&
 0.61
 \\
 060908&
 Swift&
 Long&
  19.0&
 $  9.1\pm  0.3$&
 $1.09\pm0.06$&
 0.31
 \\
 060912&
 Swift&
 Long&
   6.0&
 $ 1.53\pm 0.06$&
 $2.02\pm0.05$&
 1.57
 \\
 060927&
 Swift&
 Long&
  10.0&
 $  3.9\pm  0.1$&
 $1.92\pm0.05$&
 0.44
 \\
 061004&
 Swift&
 Long&
   6.2&
 $  2.4\pm  0.1$&
 $1.89\pm0.07$&
 0.28
 \\
 061006&
 Swift&
 Long&
   129&
 $ 0.47\pm 0.01$&
 $3.12\pm0.03$&
 1.26
 \\
 061007&
 Swift&
 Long&
  74.2&
 $ 23.2\pm  0.2$&
 $1.90\pm0.01$&
 5.04
 \\
 061110A&
 Swift&
 Long&
  44.6&
 $ 13.4\pm  0.5$&
 $1.57\pm0.07$&
 0.33
 \\
 061121&
 Swift&
 Long&
  81.3&
 $ 5.34\pm 0.06$&
 $2.58\pm0.01$&
 1.28
 \\
 061201&
 Swift&
 Short&
  0.76&
 $ 0.22\pm 0.01$&
 $1.96\pm0.07$&
 0.55
 \\
 061210&
 Swift&
 Long&
  85.3&
 $0.030\pm0.001$&
 $4.29\pm0.04$&
 13.4
 \\
 061217&
 Swift&
 Short&
  0.21&
 $ 0.15\pm 0.01$&
 $1.76\pm0.16$&
 0.08
 \\
 061222B&
 Swift&
 XRF&
  40.0&
 $ 10.7\pm  0.5$&
 $1.70\pm0.08$&
 0.43
 \\
 070110&
 Swift&
 Long&
  87.3&
 $ 18.1\pm  0.6$&
 $1.63\pm0.07$&
 0.25
 \\
 070208&
 Swift&
 Long&
  47.7&
 $  2.9\pm  0.3$&
 $1.69\pm0.17$&
 1.12
 \\
 070318&
 Swift&
 Long&
   119&
 $ 12.7\pm  0.3$&
 $1.91\pm0.04$&
 0.14
 \\
 070411&
 Swift&
 Long&
   116&
 $ 27.0\pm  0.8$&
 $1.66\pm0.05$&
 0.26
 \\
 070506&
 Swift&
 Long&
   4.3&
 $  2.1\pm  0.1$&
 $1.39\pm0.12$&
 0.55
 \\
 070508&
 Swift&
 Long&
  20.8&
 $ 6.59\pm 0.09$&
 $2.07\pm0.02$&
 1.30
 \\
 070521&
 Swift&
 Long&
  38.4&
 $ 12.2\pm  0.2$&
 $1.69\pm0.03$&
 0.42
 \\
 070529&
 Swift&
 Long&
   108&
 $  8.5\pm  0.4$&
 $1.90\pm0.09$&
 0.29
 \\
 070611&
 Swift&
 Long&
  13.2&
 $  2.6\pm  0.2$&
 $1.67\pm0.18$&
 0.28
 \\
 070612A&
 Swift&
 Long&
   359&
 $ 32.4\pm  0.9$&
 $1.91\pm0.05$&
 0.27
 \\
 070714B&
 Swift&
 Long&
  63.9&
 $ 0.48\pm 0.02$&
 $2.87\pm0.07$&
 0.55
 \\
 070721B&
 Swift&
 Long&
   345&
 $ 12.5\pm  0.4$&
 $1.87\pm0.05$&
 0.40
 \\
 070724A&
 Swift&
 Short&
  0.40&
 $ 0.11\pm 0.01$&
 $1.75\pm0.16$&
 0.66
 \\
 070802&
 Swift&
 Long&
  16.9&
 $  3.5\pm  0.3$&
 $1.74\pm0.15$&
 0.27
 \\
 070810A&
 Swift&
 XRF&
  10.9&
 $  4.1\pm  0.2$&
 $1.18\pm0.10$&
 0.46
 \\
 \enddata
 \end{deluxetable}

%% file: tab3.tex
 \begin{deluxetable}{lrrrrr}
 \tablewidth{0pt}
 \tablecaption{
 Source frame parameters;
 $z$ redshift, values from tabulations in
 Amati (2006), Ghirlanda et al. (2004) and
 http://swift.gsfc.nasa.gov/docs/swift/archive/grb\_table/.
 References for all the Swift redshifts are provided on this WWW data
 table.
 $E_{wz}$ keV,
 $Q_{pz}$ ergs keV$^{-1}$,
 $T_{Lz}$ secs,
 photon energy/peak luminosity ratio $K_{z}$.
 \label{tab3}}
 \tablehead{
 \colhead{GRB}&\colhead{$z$}&
 \colhead{$E_{wz}$}&\colhead{$Q_{pz}$}&
 \colhead{$T_{Lz}$}&\colhead{$K_{z}$}}
 \startdata
 970228&
 0.695&
 $  927\pm  320$&
 $(   0.3\pm 0.1)10^{50}$&
 $  5.7\pm  0.3$&
 $  1.21\pm  0.96$
 \\
 970508&
 0.835&
 $  906\pm  279$&
 $(   0.9\pm 0.3)10^{49}$&
 $ 2.30\pm 0.06$&
 $  1.16\pm  0.83$
 \\
 971214&
 3.420&
 $ 1981\pm  431$&
 $(   1.5\pm 0.4)10^{50}$&
 $ 2.41\pm 0.04$&
 $  1.00\pm  0.51$
 \\
 980425&
 0.009&
 $  550\pm  124$&
 $(   1.7\pm 0.5)10^{45}$&
 $  16.\pm   1.$&
 $  0.18\pm  0.10$
 \\
 980613&
 1.096&
 $  807\pm  382$&
 $(   0.7\pm 0.4)10^{49}$&
 $  5.7\pm  0.7$&
 $  1.02\pm  1.12$
 \\
 980703&
 0.966&
 $ 2061\pm  460$&
 $(   3.4\pm 0.9)10^{49}$&
 $ 10.2\pm  0.2$&
 $  0.50\pm  0.26$
 \\
 990123&
 1.600&
 $ 5736\pm  732$&
 $(   7.0\pm 1.4)10^{50}$&
 $ 8.72\pm 0.09$&
 $  0.37\pm  0.11$
 \\
 990506&
 1.307&
 $ 3000\pm  674$&
 $(   3.7\pm 1.1)10^{50}$&
 $ 6.07\pm 0.04$&
 $  0.66\pm  0.35$
 \\
 990510&
 1.619&
 $ 1539\pm  215$&
 $(   1.3\pm 0.2)10^{50}$&
 $ 2.00\pm 0.02$&
 $  1.29\pm  0.42$
 \\
 990705&
 0.843&
 $ 1484\pm  189$&
 $(   1.7\pm 0.4)10^{50}$&
 $ 14.5\pm  0.5$&
 $  0.91\pm  0.29$
 \\
 990712&
 0.430&
 $  605\pm  119$&
 $(   1.3\pm 0.7)10^{49}$&
 $  2.9\pm  0.2$&
 $  1.78\pm  0.91$
 \\
 991216&
 1.020&
 $ 2732\pm  614$&
 $(   2.4\pm 0.7)10^{50}$&
 $ 1.92\pm 0.02$&
 $  0.85\pm  0.45$
 \\
 010921&
 0.450&
 $  781\pm  173$&
 $(   1.2\pm 0.4)10^{49}$&
 $  8.5\pm  0.3$&
 $  1.08\pm  0.56$
 \\
 011121&
 0.360&
 $ 1395\pm  217$&
 $(   4.3\pm 0.9)10^{49}$&
 $ 13.9\pm  0.5$&
 $  0.72\pm  0.27$
 \\
 011211&
 2.140&
 $  733\pm  123$&
 $(   1.4\pm 0.3)10^{50}$&
 $  43.\pm   2.$&
 $  1.38\pm  0.54$
 \\
 021004&
 2.335&
 $ 1072\pm  481$&
 $(   0.4\pm 0.2)10^{50}$&
 $  3.4\pm  0.2$&
 $  1.27\pm  1.32$
 \\
 021211&
 1.010&
 $  367\pm   79$&
 $(   3.1\pm 0.8)10^{49}$&
 $  1.0\pm  0.1$&
 $  4.43\pm  2.24$
 \\
 030115A&
 2.500&
 $ 1368\pm  625$&
 $(   0.3\pm 0.2)10^{50}$&
 $  2.4\pm  0.1$&
 $  1.04\pm  1.10$
 \\
 030226&
 1.980&
 $ 1249\pm  283$&
 $(   0.9\pm 0.2)10^{50}$&
 $  9.2\pm  0.3$&
 $  1.04\pm  0.55$
 \\
 030323&
 3.370&
 $  891\pm  445$&
 $(   0.4\pm 0.3)10^{50}$&
 $  0.7\pm  0.2$&
 $  2.17\pm  2.50$
 \\
 030328&
 1.520&
 $ 1106\pm  247$&
 $(   2.6\pm 0.7)10^{50}$&
 $ 18.2\pm  0.3$&
 $  1.27\pm  0.66$
 \\
 030429&
 2.650&
 $  555\pm  168$&
 $(   0.3\pm 0.1)10^{50}$&
 $  1.6\pm  0.2$&
 $  2.67\pm  1.90$
 \\
 040924&
 0.859&
 $  995\pm  137$&
 $(   0.9\pm 0.1)10^{49}$&
 $ 0.61\pm 0.04$&
 $  1.42\pm  0.46$
 \\
 041006&
 0.716&
 $  524\pm  120$&
 $(   0.4\pm 0.1)10^{50}$&
 $  7.6\pm  0.2$&
 $  2.11\pm  1.12$
 \\
 050126&
 1.290&
 $ 1444\pm  237$&
 $(   0.7\pm 0.2)10^{49}$&
 $  3.3\pm  0.2$&
 $  0.64\pm  0.25$
 \\
 050315&
 1.949&
 $  652\pm  175$&
 $(   1.3\pm 0.4)10^{50}$&
 $  7.3\pm  0.2$&
 $  2.25\pm  1.40$
 \\
 050318&
 1.440&
 $  547\pm  118$&
 $(   2.5\pm 0.7)10^{49}$&
 $  2.2\pm  0.1$&
 $  2.40\pm  1.22$
 \\
 050319&
 3.240&
 $ 1221\pm  552$&
 $(   0.6\pm 0.3)10^{50}$&
 $ 0.94\pm 0.06$&
 $  1.63\pm  1.71$
 \\
 050401&
 2.900&
 $ 2008\pm  315$&
 $(   1.9\pm 0.3)10^{50}$&
 $ 1.56\pm 0.05$&
 $  1.14\pm  0.41$
 \\
 050505&
 4.270&
 $ 2002\pm  200$&
 $(   1.0\pm 0.2)10^{50}$&
 $ 2.16\pm 0.09$&
 $  0.92\pm  0.23$
 \\
 050509B&
 0.225&
 $  738\pm  220$&
 $(   0.4\pm 0.2)10^{46}$&
 $0.035\pm0.004$&
 $  0.65\pm  0.46$
 \\
 050525A&
 0.606&
 $  537\pm   59$&
 $(   6.2\pm 0.8)10^{49}$&
 $ 2.46\pm 0.07$&
 $  2.91\pm  0.75$
 \\
 050603&
 2.821&
 $ 4123\pm  528$&
 $(   1.2\pm 0.2)10^{50}$&
 $0.219\pm0.005$&
 $  0.79\pm  0.23$
 \\
 050709&
 0.160&
 $  379\pm   59$&
 $(   1.0\pm 0.2)10^{47}$&
 $0.048\pm0.003$&
 $  2.33\pm  0.86$
 \\
 050724&
 0.257&
 $  229\pm  114$&
 $(   0.2\pm 0.2)10^{49}$&
 $ 0.38\pm 0.02$&
 $  4.72\pm  5.43$
 \\
 050730&
 3.970&
 $ 2454\pm  378$&
 $(   0.7\pm 0.1)10^{50}$&
 $  5.2\pm  0.2$&
 $  0.58\pm  0.21$
 \\
 050802&
 1.710&
 $ 1423\pm  219$&
 $(   2.6\pm 0.5)10^{49}$&
 $  2.3\pm  0.1$&
 $  0.95\pm  0.34$
 \\
 050803&
 0.422&
 $  842\pm  110$&
 $(   3.1\pm 0.5)10^{48}$&
 $ 10.1\pm  0.3$&
 $  0.71\pm  0.22$
 \\
 050813&
 1.800&
 $ 1631\pm  461$&
 $(   0.6\pm 0.2)10^{48}$&
 $0.026\pm0.003$&
 $  0.96\pm  0.65$
 \\
 050814&
 5.300&
 $ 1960\pm  682$&
 $(   1.3\pm 0.5)10^{50}$&
 $ 1.76\pm 0.09$&
 $  1.04\pm  0.84$
 \\
 050820A&
 2.612&
 $ 3082\pm 1241$&
 $(   0.6\pm 0.2)10^{50}$&
 $ 3.20\pm 0.07$&
 $  0.49\pm  0.46$
 \\
 050904&
 6.290&
 $ 8600\pm 3034$&
 $(   1.3\pm 0.5)10^{50}$&
 $  7.8\pm  0.2$&
 $  0.18\pm  0.15$
 \\
 050908&
 3.350&
 $ 1609\pm  418$&
 $(   1.7\pm 0.5)10^{49}$&
 $ 1.27\pm 0.06$&
 $  0.87\pm  0.53$
 \\
 050922C&
 2.198&
 $ 1618\pm  488$&
 $(   2.9\pm 0.9)10^{49}$&
 $ 0.57\pm 0.02$&
 $  1.16\pm  0.81$
 \\
 051022&
 0.800&
 $ 3431\pm  416$&
 $(   1.6\pm 0.2)10^{50}$&
 $  6.0\pm  0.2$&
 $  0.48\pm  0.14$
 \\
 051109A&
 2.346&
 $ 2149\pm 1074$&
 $(   0.4\pm 0.3)10^{50}$&
 $ 0.85\pm 0.07$&
 $  0.84\pm  0.97$
 \\
 051109B&
 0.080&
 $  192\pm   96$&
 $(   0.4\pm 0.4)10^{47}$&
 $  1.8\pm  0.2$&
 $  1.71\pm  1.97$
 \\
 051111&
 1.549&
 $ 2036\pm  523$&
 $(   0.4\pm 0.1)10^{50}$&
 $  5.2\pm  0.1$&
 $  0.61\pm  0.36$
 \\
 051221A&
 0.547&
 $ 2349\pm  481$&
 $(   1.8\pm 0.4)10^{48}$&
 $0.077\pm0.002$&
 $  0.68\pm  0.32$
 \\
 060108&
 2.030&
 $  896\pm  320$&
 $(   1.1\pm 0.4)10^{49}$&
 $ 1.44\pm 0.08$&
 $  1.36\pm  1.12$
 \\
 060115&
 3.530&
 $ 1152\pm  353$&
 $(   0.8\pm 0.3)10^{50}$&
 $  3.9\pm  0.1$&
 $  1.33\pm  0.95$
 \\
 060116&
 6.600&
 $ 3865\pm  660$&
 $(   1.0\pm 0.2)10^{50}$&
 $ 1.20\pm 0.07$&
 $  0.55\pm  0.22$
 \\
 060124&
 2.297&
 $ 3439\pm  938$&
 $(   0.5\pm 0.2)10^{49}$&
 $ 1.24\pm 0.08$&
 $  0.32\pm  0.20$
 \\
 060206&
 4.050&
 $ 1728\pm  533$&
 $(   0.4\pm 0.1)10^{50}$&
 $ 0.65\pm 0.02$&
 $  1.11\pm  0.80$
 \\
 060210&
 3.910&
 $ 2381\pm  422$&
 $(   2.3\pm 0.5)10^{50}$&
 $  5.1\pm  0.1$&
 $  0.78\pm  0.32$
 \\
 060218&
 0.030&
 $  123\pm   61$&
 $(   0.6\pm 0.4)10^{47}$&
 $ 19.8\pm  0.9$&
 $  1.65\pm  1.90$
 \\
 060223A&
 4.410&
 $ 2108\pm  399$&
 $(   2.8\pm 0.6)10^{49}$&
 $ 0.86\pm 0.05$&
 $  0.81\pm  0.36$
 \\
 060418&
 1.489&
 $ 2509\pm  561$&
 $(   0.6\pm 0.1)10^{50}$&
 $  6.6\pm  0.1$&
 $  0.51\pm  0.26$
 \\
 060502A&
 1.510&
 $ 1399\pm  158$&
 $(   2.5\pm 0.3)10^{49}$&
 $  4.8\pm  0.2$&
 $  0.81\pm  0.21$
 \\
 060502B&
 0.287&
 $  833\pm  217$&
 $(   0.3\pm 0.1)10^{47}$&
 $0.045\pm0.004$&
 $  0.84\pm  0.52$
 \\
 060510B&
 4.900&
 $ 2477\pm  308$&
 $(   1.7\pm 0.4)10^{50}$&
 $  9.6\pm  0.2$&
 $  0.61\pm  0.19$
 \\
 060522&
 5.110&
 $ 2832\pm  515$&
 $(   0.4\pm 0.1)10^{50}$&
 $ 1.49\pm 0.07$&
 $  0.59\pm  0.25$
 \\
 060526&
 3.210&
 $  995\pm  497$&
 $(   0.8\pm 0.5)10^{50}$&
 $ 1.58\pm 0.08$&
 $  1.86\pm  2.14$
 \\
 060604&
 2.680&
 $ 1209\pm  604$&
 $(   0.1\pm 0.1)10^{50}$&
 $ 0.62\pm 0.06$&
 $  1.29\pm  1.48$
 \\
 060605&
 3.800&
 $ 2692\pm  386$&
 $(   2.0\pm 0.4)10^{49}$&
 $ 1.64\pm 0.07$&
 $  0.51\pm  0.18$
 \\
 060607A&
 3.082&
 $ 2033\pm  653$&
 $(   0.6\pm 0.2)10^{50}$&
 $ 3.73\pm 0.10$&
 $  0.73\pm  0.54$
 \\
 060614&
 0.125&
 $  434\pm  217$&
 $(   0.5\pm 0.3)10^{49}$&
 $ 23.7\pm  0.3$&
 $  1.22\pm  1.40$
 \\
 060707&
 3.430&
 $  975\pm  487$&
 $(   0.8\pm 0.7)10^{50}$&
 $  2.3\pm  0.1$&
 $  1.76\pm  2.03$
 \\
 060714&
 2.710&
 $ 1115\pm  271$&
 $(   1.1\pm 0.3)10^{50}$&
 $  6.2\pm  0.2$&
 $  1.33\pm  0.75$
 \\
 060729&
 0.540&
 $  587\pm  144$&
 $(   0.8\pm 0.2)10^{49}$&
 $ 11.2\pm  0.3$&
 $  1.23\pm  0.70$
 \\
 060801&
 1.130&
 $ 1918\pm  265$&
 $(   0.9\pm 0.2)10^{48}$&
 $0.095\pm0.006$&
 $  0.68\pm  0.23$
 \\
 060814&
 0.840&
 $ 2172\pm  791$&
 $(   0.4\pm 0.2)10^{50}$&
 $ 13.9\pm  0.1$&
 $  0.46\pm  0.39$
 \\
 060904B&
 0.703&
 $  742\pm  155$&
 $(   0.7\pm 0.2)10^{49}$&
 $ 2.88\pm 0.09$&
 $  1.27\pm  0.62$
 \\
 060906&
 3.685&
 $ 1202\pm  352$&
 $(   1.4\pm 0.4)10^{50}$&
 $ 2.59\pm 0.09$&
 $  1.58\pm  1.07$
 \\
 060908&
 2.430&
 $ 1882\pm  403$&
 $(   0.5\pm 0.1)10^{50}$&
 $ 2.66\pm 0.08$&
 $  0.82\pm  0.41$
 \\
 060912&
 0.937&
 $  792\pm  117$&
 $(   1.0\pm 0.2)10^{49}$&
 $ 0.79\pm 0.03$&
 $  1.70\pm  0.59$
 \\
 060927&
 5.600&
 $ 1761\pm  474$&
 $(   0.7\pm 0.2)10^{50}$&
 $ 0.59\pm 0.02$&
 $  1.28\pm  0.81$
 \\
 061004&
 3.300&
 $ 1668\pm  273$&
 $(   1.9\pm 0.4)10^{49}$&
 $ 0.55\pm 0.03$&
 $  1.03\pm  0.39$
 \\
 061006&
 0.438&
 $ 2981\pm  886$&
 $(   2.5\pm 0.9)10^{48}$&
 $0.328\pm0.008$&
 $  0.42\pm  0.29$
 \\
 061007&
 1.261&
 $ 2891\pm  319$&
 $(   4.0\pm 0.4)10^{50}$&
 $10.28\pm 0.08$&
 $  0.62\pm  0.16$
 \\
 061110A&
 0.758&
 $  806\pm  137$&
 $(   0.5\pm 0.1)10^{49}$&
 $  7.6\pm  0.3$&
 $  0.87\pm  0.35$
 \\
 061121&
 1.314&
 $ 4343\pm  673$&
 $(   0.8\pm 0.1)10^{50}$&
 $ 2.31\pm 0.03$&
 $  0.41\pm  0.15$
 \\
 061201&
 0.835&
 $ 4280\pm 1868$&
 $(   0.3\pm 0.1)10^{49}$&
 $0.118\pm0.005$&
 $  0.38\pm  0.38$
 \\
 061210&
 0.410&
 $  755\pm  209$&
 $(   1.6\pm 0.6)10^{48}$&
 $0.021\pm0.001$&
 $  2.67\pm  1.75$
 \\
 061217&
 0.287&
 $ 1082\pm  182$&
 $(   0.4\pm 0.1)10^{47}$&
 $ 0.12\pm 0.01$&
 $  0.56\pm  0.23$
 \\
 061222B&
 3.360&
 $ 1030\pm  115$&
 $(   1.4\pm 0.4)10^{50}$&
 $  2.4\pm  0.1$&
 $  1.87\pm  0.52$
 \\
 070110&
 2.352&
 $ 1618\pm  234$&
 $(   3.2\pm 0.6)10^{49}$&
 $  5.4\pm  0.2$&
 $  0.72\pm  0.24$
 \\
 070208&
 1.170&
 $  830\pm  415$&
 $(   0.5\pm 0.3)10^{49}$&
 $  1.3\pm  0.1$&
 $  1.24\pm  1.43$
 \\
 070318&
 0.840&
 $ 1054\pm  125$&
 $(   1.2\pm 0.2)10^{49}$&
 $  6.9\pm  0.2$&
 $  0.83\pm  0.23$
 \\
 070411&
 2.950&
 $ 1718\pm  252$&
 $(   0.7\pm 0.1)10^{50}$&
 $  6.8\pm  0.2$&
 $  0.77\pm  0.26$
 \\
 070506&
 2.310&
 $  712\pm  356$&
 $(   0.8\pm 0.5)10^{49}$&
 $ 0.65\pm 0.05$&
 $  1.88\pm  2.16$
 \\
 070508&
 0.820&
 $ 1177\pm  119$&
 $(   8.7\pm 0.9)10^{49}$&
 $ 3.62\pm 0.05$&
 $  1.34\pm  0.31$
 \\
 070521&
 0.553&
 $ 1007\pm  104$&
 $(   1.8\pm 0.2)10^{49}$&
 $  7.8\pm  0.2$&
 $  0.94\pm  0.22$
 \\
 070529&
 2.500&
 $ 1841\pm  309$&
 $(   0.5\pm 0.1)10^{50}$&
 $  2.4\pm  0.1$&
 $  0.84\pm  0.33$
 \\
 070611&
 2.040&
 $ 1231\pm  414$&
 $(   0.8\pm 0.3)10^{49}$&
 $ 0.86\pm 0.08$&
 $  1.04\pm  0.81$
 \\
 070612A&
 0.617&
 $  699\pm  108$&
 $(   3.7\pm 0.7)10^{49}$&
 $ 20.0\pm  0.6$&
 $  1.27\pm  0.46$
 \\
 070714B&
 0.920&
 $ 1117\pm  207$&
 $(   3.4\pm 0.9)10^{48}$&
 $ 0.25\pm 0.01$&
 $  1.25\pm  0.55$
 \\
 070721B&
 3.626&
 $ 2698\pm  329$&
 $(   0.8\pm 0.1)10^{50}$&
 $ 2.71\pm 0.08$&
 $  0.63\pm  0.18$
 \\
 070724A&
 0.457&
 $  556\pm  278$&
 $(   0.7\pm 0.4)10^{47}$&
 $0.074\pm0.007$&
 $  1.35\pm  1.55$
 \\
 070802&
 2.450&
 $ 1417\pm  462$&
 $(   0.6\pm 0.3)10^{49}$&
 $ 1.02\pm 0.08$&
 $  0.82\pm  0.63$
 \\
 070810A&
 2.170&
 $  550\pm  275$&
 $(   0.4\pm 0.2)10^{50}$&
 $ 1.29\pm 0.07$&
 $  2.95\pm  3.40$
 \\
 \enddata
 \end{deluxetable}